\documentclass[10pt, onecolumn]{IEEEtran}

\usepackage{graphicx}
\usepackage{xr-hyper}
\usepackage[ruled,vlined]{algorithm2e}
\usepackage{subfigure}
\usepackage{multirow}
\usepackage{enumerate}
\usepackage{pifont}
\usepackage{array}
\usepackage{graphicx}
\usepackage{array}
\usepackage{tabulary}
\usepackage{capt-of}

\begin{document}

\title{mPDF: Framework for Watermarking PDF Files using Image Watermarking Algorithms}

\author{
Sachin~Mehta*,~Balakrishnan~Prabhakaran, ~Rajarathnam~Nallusamy, ~and~ Derrick Newton
\thanks{Sachin Mehta is a graduate student at Department of Electrical Engineering, University of Washington, Seattle, USA. (e-mail: sacmehta@uw.edu)}
\thanks{Balakrishnan~Prabhakaran is a Professor with Department of Computer Science, University of Texas at Dallas, Richardson, TX75080 USA (email:bprabhakaran@utdallas.edu)}
\thanks{Rajarathnam~Nallusamy is a Director at Embnology Solutions Private Limited, Bangalore - 560037 (e-mail: raja@embnology.com)}
\thanks{Derrick Newton is a Lecturer at Faculty of Engineering and Computing, Coventry University, UK (email:ab3729@coventry.ac.uk)}
\thanks{This work was done as a part of when Sachin Mehta and Rajarathnam Nallusamy were working with Infosys Research Lab, Bangalore, India - 560100.}
}

\maketitle

\begin{abstract}
The advancement in digital technologies have made it possible to produce perfect copies of digital content. In this environment, malicious users reproduce the digital content and share it without compensation to the content owner. Content owners are concerned about the potential loss of revenue and reputation from piracy, especially when the content is available over the Internet. Digital watermarking has emerged as a deterrent measure towards such malicious activities. Several methods have been proposed for copyright protection and fingerprinting of digital images. However, these methods are not applicable to text documents as these documents lack rich texture information which is abundantly available in digital images. In this paper, a  framework (mPDF) is proposed which facilitates the usage of digital image watermarking algorithms on text documents. The proposed method divides a text document into texture and non-texture blocks using an energy-based approach. After classification, a watermark is embedded inside the texture blocks in a content adaptive manner. The proposed method is integrated with five known image watermarking methods and its performance is studied in terms of quality and robustness. Experiments are conducted on documents in 11 different languages. Experimental results clearly show that the proposed method facilitates the usage of image watermarking algorithms on text documents and is robust against attacks such as print \& scan, print screen, and skew. Also, the proposed method overcomes the drawbacks of existing text watermarking methods such as manual inspection and language dependency.
\end{abstract}


\maketitle

\section{Introduction}
The advancement in digital technology has revolutionized traditional business models. Nowadays, consumers can create, share, and distribute digital content across the world in fractions of a second. On one hand, these advancements have made human life easy, but on the other hand, it has increased piracy rates. Pirated digital content is readily available over the Internet.\\
%%%%
Digital Rights Management (DRM) based systems are available for protecting content from misuse. These systems restrict access to the content and protect the document from misuse until it is encrypted or access restricted \cite{Zeng:eke}. Once the document is obtained in clear text form, these systems might not be able to protect digital rights \cite{HernandezArdieta:eke}. A simple example for creating a nearly similar copy of content is using \textit{print screen}. A malicious user can take snapshots of all the pages of the document and construct a nearly similar copy which can be distributed illegally. Some DRM based systems disable such options while the content is being viewed, but these systems are very expensive \cite{CopySafe2014:eke}. Also, these systems are difficult to implement in an organization having heterogeneous networks. Apart from their own network, these organizations have other networks, usually of their clients. In such a case, organizations fail to use the features of DRM solutions and fail to protect the content.

\textbf{Motivation and Contributions}: Digital watermarking has emerged as a deterrent measure to protect the digital rights. Researchers have proposed several methods for copyright protection and traitor tracing of digital images. However, the application of these methods on text documents poses different challenges. 

\begin{figure}[t!]
\centering
\includegraphics[width=9cm]{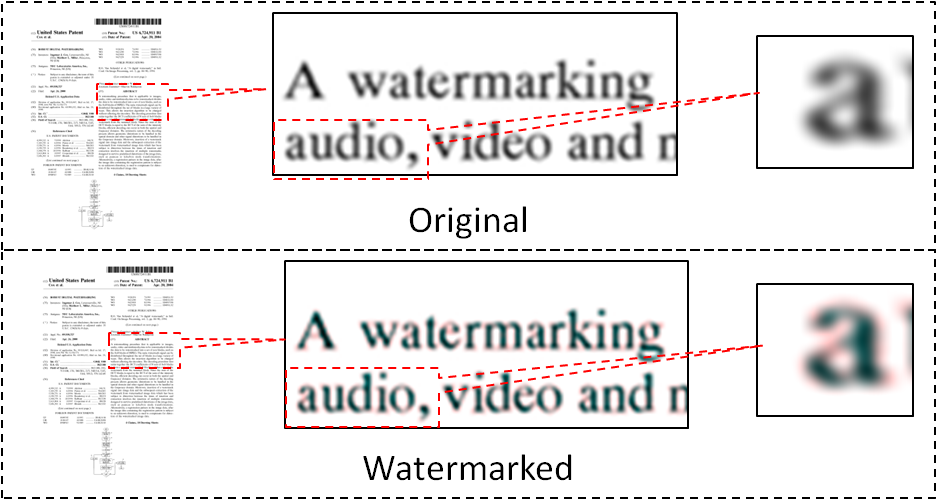}
\caption{Impact of image watermarking algorithms on text documents}
\label{fig:distortion}
\end{figure}

Digital watermarking aims at changing the pixel values such that changes are unnoticeable under normal viewing conditions. Pixels in images (either grey scale or color) take values from a wide range. For most pixels, the distortions resulting due to watermarking remain unnoticeable. Unlike images, pixels in text documents take values from a few possibilities. Due to lack of rich texture information in text documents, hiding data using image watermarking algorithms becomes difficult \cite{minWuAndLiu:eke}. In such cases, watermark may be either lost or lead to visible distortions, as shown in Figure \ref{fig:distortion}. To avoid such scenarios, researchers have proposed different methods for text watermarking and these are discussed in Section \ref{sec:realted}. These methods suffer from several drawbacks such as limited applications and manual introspection.
%===============

\begin{figure*}[t!]
\centering
\includegraphics[width=10cm]{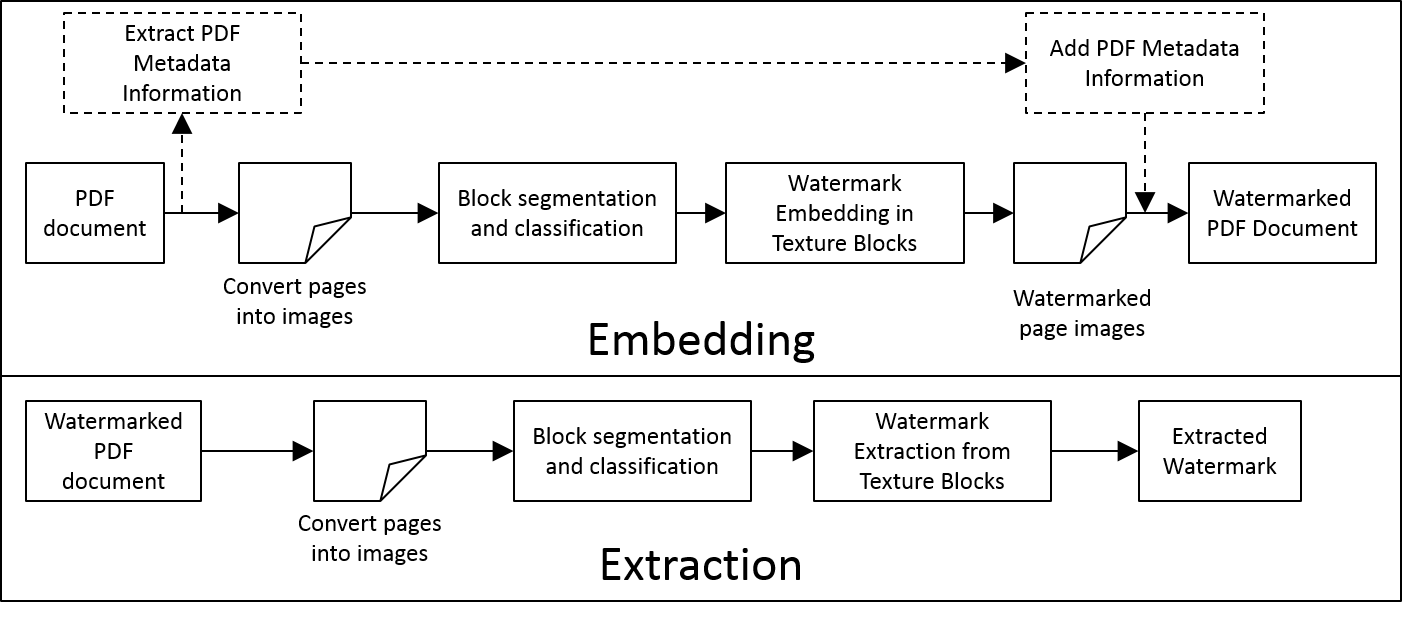}
\caption{Overview of the proposed method}
\label{fig:overview}
\end{figure*}

In this paper, we propose an approach which overcomes the drawbacks of existing methods. Figure \ref{fig:overview} provides an overview of the proposed method. We propose a framework which facilitates the usage of known image watermarking methods on text documents. To the best of our knowledge, there has been no work on developing techniques to watermark the text documents using image watermarking algorithms. In particular, we propose an energy-based block classification approach which classifies the blocks into different categories (texture and non-texture blocks). A watermark is then embedded inside texture blocks using a content adaptive watermark embedding strength. We created a data set of text documents from 11 different languages and studied the imperceptibility, capacity, and robustness of 5 known image watermarking algorithms.
 
Rest of this paper is organized as: Related work is discussed in Section \ref{sec:realted}. The proposed method is discussed in Section \ref{sec:proposedMethod}. Criteria to evaluate the performance of the proposed method is given in Section \ref{sec:criteria}. Experimental results are discussed in Section \ref{sec:results}. Conclusions are drawn in Section \ref{sec:conclusion}.
%%%%%%%%
\section{Related Work}
\label{sec:realted}
Text watermarking methods can be classified into the following categories:
\begin{enumerate}
\item \textbf{Character Feature Methods: } These methods manipulate the features of characters such as shape, size, and position. Wu et al. \cite{minWuAndLiu:eke} proposed a flippable pixels based data hiding method. The method reads each page of the document as an image and determines the flippable pixels using a set of rules. These flippable pixels are then manipulated for embedding the watermark. Wenyin and Ningde \cite{Wenyin2006:eke} proposed a watermarking method in which specific characteristics of Chinese characters are used for embedding the watermark. The method detects Chinese characters with occlusive components (characters with one or more hollow closing regions) and uses these characters for embedding the watermark. A color quantization based watermarking method was proposed in \cite{Borges2008:eke} \cite{Villan:eke}. Based on the watermark bit, the method changes the color of the character.\\
Character feature methods, in general, exploit the characteristics of the alphabets in any language for embedding the watermark, for example occlusive components in Chinese alphabets. These methods are language specific and have applications limited to tamper detection and document identification. If a different watermark is embedded using these methods for fingerprinting, then a malicious user can easily detect or destroy the watermark by a simple comparative operation between two different copies of same content. 
\item \textbf{Open Space Methods:} These methods embed the watermark by modulating either inter-line or inter-word or inter-character space. Huang et al.  \cite{Huang2001:eke} and Alattar et al. \cite{Adnan2004:eke} proposed a text watermarking method in which inter-word spaces and inter-line distances are modified for embedding the watermark. \\
These methods have high embedding capacity. Like character feature methods, these methods have limited applications and cannot be used in fingerprinting.
\item \textbf{Zero Watermarking Methods:} Instead of embedding a watermark, these methods construct a watermark using  text document features such as occurrence of words with four or more characters. Zhang et al. \cite{Jixian2012:eke} proposed a text watermarking algorithm in which features such as high frequency words and sentence length are used to create a watermark. Yawai and Hiransakolwong \cite{Yawai2012:eke} proposed a zero watermarking method using line intersection. The method determines the intersection point for each character which is then used as a watermarking point. \\
These methods construct the watermark from the document itself and hence, cannot be used for fingerprinting applications. 
\item \textbf{Natural Language Watermarking (NLW):} These methods replace the words by their synonyms or sentences are transformed via suppression or inclusion of noun phrases. Topkara et al. \cite{Topkara2006:eke} proposed a NLW method. To embed the watermark, the method transforms the sentence. Halvani et al. \cite{Halvani2013:eke} proposed a NLW method for German language. The method embeds the watermark by lexical and syntactic substitution.\\
NLW methods are syntax and semantic based. A manual inspection is required for validating the accuracy of these methods. Further, these methods are difficult to extend for fingerprinting applications. 
\item \textbf{Visible Watermarking:} These methods embed the visible watermark. Microsoft's word also offers this feature. As the watermark is visible, the consumers of the document can be easily identified. However, these methods cannot be used for forensic applications as a malicious user may try to frame an innocent user.
\end{enumerate}

Existing methods suffer from following drawbacks:
\begin{itemize}
\item \textit{Limited Applications:} These methods can be used in authentication, tamper detection, and copyright protection. These methods are not suitable for fingerprinting which aims at embedding a unique watermark for every consumer.
\item \textit{Multi-lingual Support:} Most of these methods utilize the specific characteristics of a particular
language for watermarking which make their application to other language documents very difficult.
\item \textit{Manual Introspection:} NLW methods are based on substitution either at sentence or word level. Sometimes substitution may lead to a change in the meaning of the sentence. Hence, every watermarked document needs to be manually inspected. This is a tedious process and make the method practically infeasible.
\end{itemize}
This paper proposes a method which addresses the drawbacks of the existing methods such as limited applications and language dependency.
%%%%%%%%%%%%%%%%%%%%%%%%
\section{The Proposed Method}
\label{sec:proposedMethod}
For addressing the challenges involved in adopting image watermarking strategies for text documents, we propose two main techniques: (i) content-based block classification; (ii) determination of content adaptive watermarking strength.
\subsection{Content based block classification}
\label{ssec:contentClassify}
The proposed method segments the page into blocks and selects a few blocks using the selection protocol to embed the watermark.
\subsubsection{Block Analysis}
\label{sssec:blockAnalysis}
\begin{figure}[t!]
\centering
\subfigure[]{\makebox[3 cm][c]{\fbox{\includegraphics[scale=0.25]{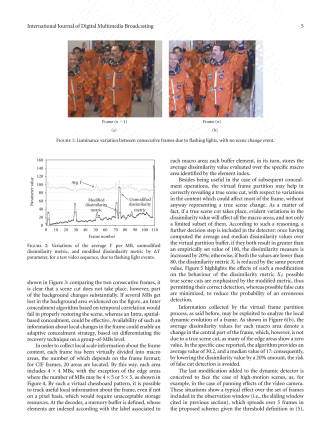}}}}%
\hspace{0.001cm}
\subfigure[]{\makebox[3 cm][c]{\includegraphics[scale=0.25]{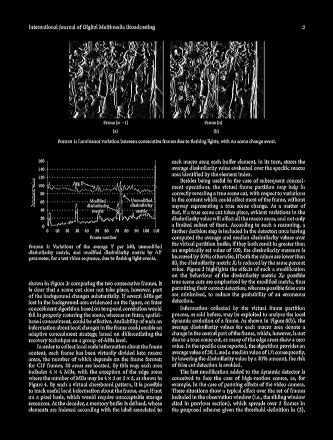}}}%
\hspace{0.001cm}
\subfigure[]{\makebox[3 cm][c]{\includegraphics[scale=0.25]{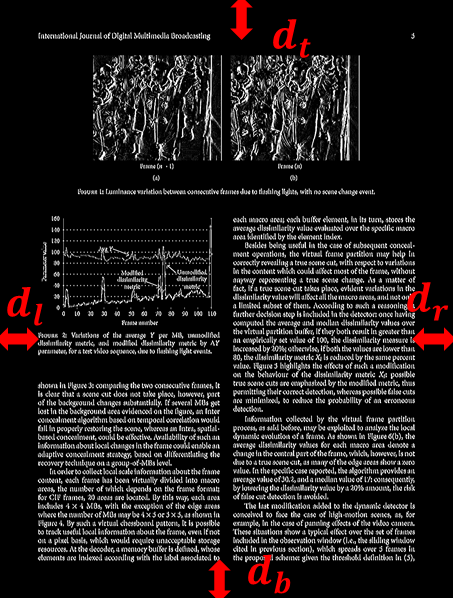}}}%
\hspace{0.001cm}
\subfigure[]{\makebox[3 cm][c]{\fbox{\includegraphics[scale=0.25]{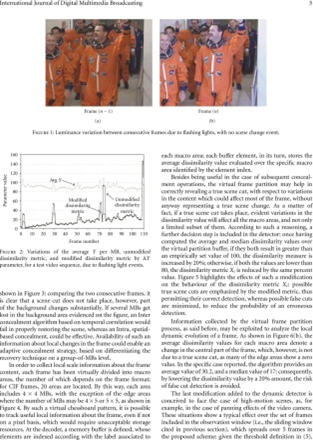}}}}%
\caption{Differentiation operation and cropping: (a) Original Page $P$, (b) Output of SOBEL operator, (c) Distances from different directions, and (d) Cropped page $P_C$}
\label{fig:outDiffCrop}
\end{figure}

\begin{figure}[t!]
\centering
\subfigure[CW]{\makebox[2cm]{\fbox{\includegraphics[scale=0.6]{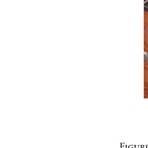}}}}
\subfigure[PT]{\makebox[2cm]{\fbox{\includegraphics[scale=0.6]{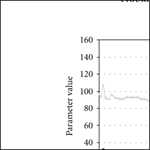}}}}
\subfigure[CT]{\makebox[2cm]{\fbox{\includegraphics[scale=0.6]{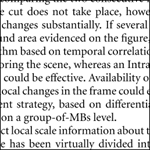}}}}
\subfigure[PTPG]{\makebox[2cm]{\fbox{\includegraphics[scale=0.6]{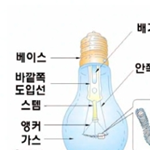}}}}
\subfigure[PTPG]{\makebox[2cm]{\fbox{\includegraphics[scale=0.6]{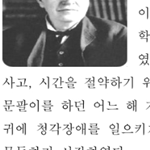}}}}
\caption{Different types of blocks: (a -b) Non-texture blocks and (c-e) Texture blocks}
\label{fig:differentBlocks}
\end{figure}

\begin{figure}[t!]
\begin{minipage}{.5\textwidth}
\centering
\includegraphics[width=0.85\columnwidth]{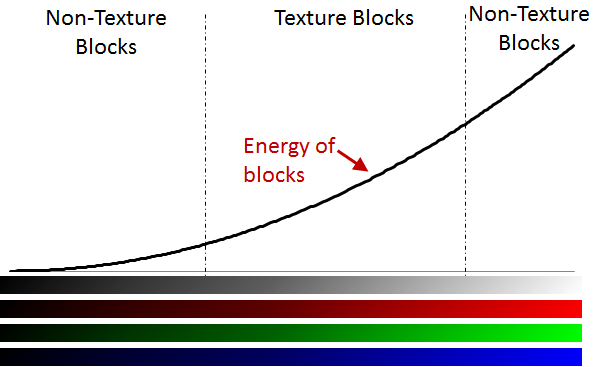}
\caption[Energy based classification of blocks]{Energy based classification of blocks [x-axis denotes the pixel values of the block while y-axis denotes the energy of the block.]}
\label{fig:energyBasedSegmentation}
\end{minipage}
\hspace{0.5cm}
\begin{minipage}{.5\textwidth}
\centering
\fbox{\includegraphics[width=0.85\columnwidth]{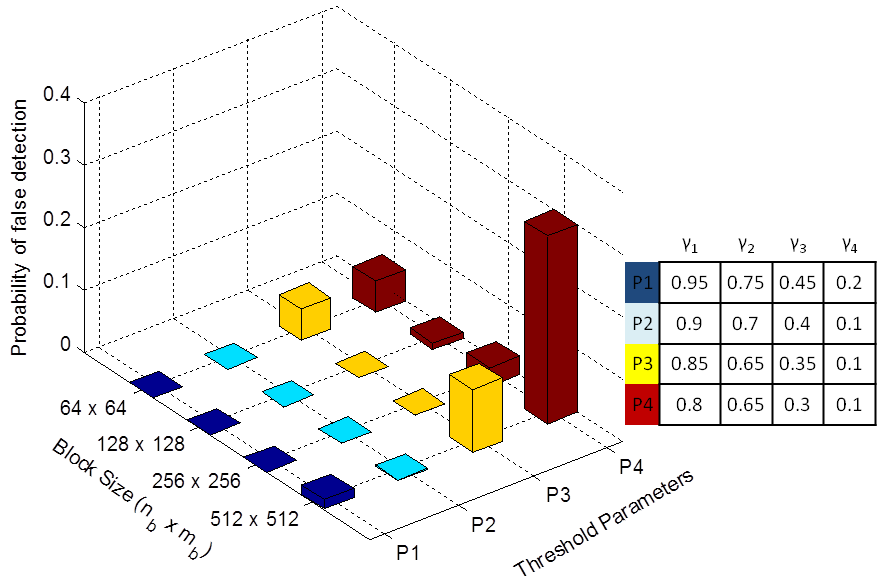}}
\caption{Relationship between block classification, block size, and thresholds}
\label{fig:blockProb}
\end{minipage}
\end{figure}

Text document contains margins. These margins do not contain any information. The proposed method determine these margins before segmenting the document into blocks. Once the information about margins is obtained, the page is cropped and then, segmented into blocks which are classified using the selection protocol.
\begin{itemize}
\itemsep 0em
\item \textbf{Detection of Page Margins and Cropping:} Let us assume that page $P$ of a text document has dimension $N \times M$. Margins in a document can be easily detected in a text document. A simple way to detect the margin is to identify a first \emph{non-white} pixel from all the sides of page. The distance of the first non-white pixel from the document edge denotes the margin. However, this simple approach fails when some distortions are present in the document. Intentional or un-intentional modification of the text document may lead to change in certain pixel values, say in compression. These changes will lead to false identification of margins. Figure \ref{fig:exampleFalseMargin} shows an example of false margin detection. Difference in  dimension of the cropped page with and without distortion clearly indicates that the simple approach fails to detect margins correctly when distortions are present.\\
%%%%%%%%%%%%%%%
\begin{figure}[b!]
\centering
\subfigure[$1361\times 1952$]{\includegraphics[width=3.5cm]{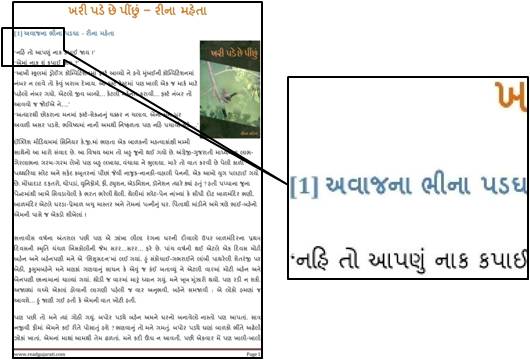}}%
\hspace{1cm}
\subfigure[$1403\times 1975$]{\includegraphics[width=3.5cm]{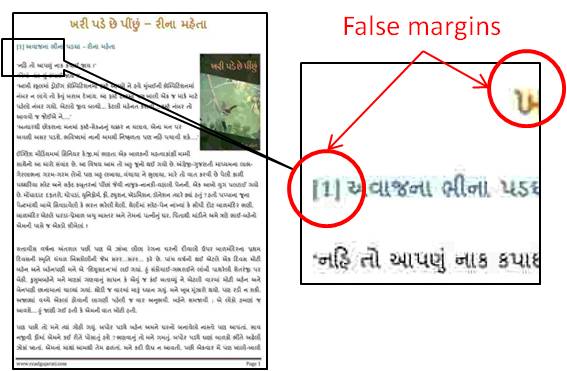}}%
\caption{False margin detection: (a) Cropped page without compression (b) Cropped page with compression (Quality factor=10)}
\label{fig:exampleFalseMargin}
\end{figure}
%%%%%%%%%%
To overcome such problems, we pre-process the page before identification of margins. The proposed method applies a discrete differentiation operator on the page. The proposed method uses a SOBEL operator for differentiation. For page $P$ with coordinates $(x,y)$, the magnitude of the gradient can be expressed as:
\begin{equation}
mag(\bigtriangledown P) = \sqrt{|G_x^2\ +\ G_y^2|},\ where\ \ 
G_y = \left[ 
\begin{array}{ccc}
 -1 & -2 & -1 \\
0 & 0 & 0\\
1 & 2 & 1\\
\end{array}
\right]
* P, \ \ 
G_x = \left[ 
\begin{array}{ccc}
 -1 & 0 & 1 \\
-2 & 0 & 2\\
-1 & 2 & 1\\
\end{array}
\right]
* P
\label{eq:gradient}
\end{equation}
Once output of differentiation operator is obtained, distance of first \emph{white pixel} is computed from all directions\footnote{The proposed method doesn't apply any thresholding to the output of the differentiation operator}. These distances can be represented as $d_l$, $d_r$, $d_t$, and $d_b$ for left, right, top, and bottom sides respectively. Once distances are determined, page $P$ is cropped to obtain the cropped page $P_C$ having dimension $N_C \times M_C$. An example of the differentiation operator and cropping is shown in Figure \ref{fig:outDiffCrop}.
\item \textbf{Segmentation and Classification of Blocks:} Once the cropped page $P_C$ is obtained, it is segmented into blocks of dimension $n \times m$. Let us say that $b$ such blocks exist such that $P_C = \{B(1), B(2), ..., B(b)\}$
where $B(\cdot)$ represents the block. These $b$ blocks are classified into two categories: (i) texture blocks and (ii) non-texture blocks. Texture blocks are the ones which contain either complete text (CT) or complete graphics (CG) or partial text and partial graphics (PTPG). Non-texture blocks are either completely white (CW) or completely black (CB) or blocks with partial texts (PT).  Examples of different types of blocks are shown in Figure \ref{fig:differentBlocks}. The proposed method classify the blocks using energy E of the block.  Figure \ref{fig:energyBasedSegmentation} shows the classification of blocks using energy in gray level and RGB color space \footnote{For illustration purpose, we have assumed that all pixels in a block contain same values.}.

The proposed method uses DCT to compute the energy of a block. This choice is made due to the \textit{energy compaction} property of DCT where in the energy of the entire block is concentrated in a DC component and a few AC components. Blocks in text documents do not have rich texture information and hence, the majority of the energy of the block lies in DC coefficient. The proposed method classifies the block using the DC coefficient and the classification criteria is explained in Algorithm \ref{algo:contentClassi}. The performance of Algorithm \ref{algo:contentClassi} is dependent upon the thresholds. Figure \ref{fig:blockProb} shows the relation between the probability of false classification, block size, and thresholds. From Figure \ref{fig:blockProb}, it is clear that probability of false classification of blocks is less when $\gamma_1 = 0.9$, $\gamma_2 = 0.7$,$\gamma_3 = 0.4$, and $\gamma_4 = 0.1$. We have used the same values in our experiments \footnote{Histogram-based approach can classify the blocks only into two categories i.e. text and graphic blocks and is discussed in Appendix \ref{sec:histApproach}.}.
%================================================
\end{itemize}
%===
%========
%=====================
%========

\begin{algorithm}[H]
\scriptsize
%\small
 \caption{Energy-based classification of Blocks}
 $B_{DCT} = dct(B(i)), \quad 1\leq i \leq b$
\BlankLine
 $DC = B_{DCT}(0,0)$
 \BlankLine
 \uIf {$DC > T_1$ or $DC < T_4$}{
   \emph{Non-texture block}
 }
 \uElse{
    \uIf {$DC \leq T_1$ and $DC > T_2$}{
       \emph{Complete text block}
    }
    \uElseIf {$DC \leq T_2$ and $DC > T_3$}{
       \emph{Partial graphics and partial text}
    }
    \uElseIf {$DC \leq T_3$ and $DC > T_4$}{
       \emph{Complete Graphics}
    }
 }
\BlankLine
where\\
\BlankLine
$T_1$, $T_2$, $T_3$, and $T_4$ are content thresholds used for block classification.
\BlankLine
$T_1 = \gamma_1 \times B_{max}$, $T_2 = \gamma_2 \times B_{max}$, $T_3 = \gamma_3 \times B_{max}$, and $T_4 = \gamma_4 \times B_{max}$
\BlankLine
$B_{max} = \sqrt{\frac{\sum_{i=1}^{n}\sum_{j=1}^{m} 255^2}{n \times m}} =$ maximum possible energy of the block
\BlankLine
$\gamma_1$, $\gamma_2$, $\gamma_3$, and $\gamma_4$ are the constants used for setting the threshold values.
\label{algo:contentClassi}
\end{algorithm}
%=============================================
%===================================
%%%
\subsection{Watermarking}
\label{ssec:watermarking}
\begin{itemize}
\itemsep 0em
\item \textbf{Content adaptive watermark embedding strength:} Once the blocks are classified, we need to embed the watermark inside these blocks. Fixed watermarking strengths may either lead to perceptibility (as shown in Figure \ref{fig:distortion}) or automatic destruction of the watermark, as in completely white blocks where all the pixels in RGB color space have value (255, 255, 255). Watermark is like a noise and it's addition will increase the pixel value from (255, 255, 255) to (255 + $\Delta$, 255 + $\Delta$, 255 + $\Delta$) assuming watermark is added uniformly across RGB channels. Here, $\Delta$ represents the change in pixel value due to watermark addition. Since a pixel can have a maximum value of (255, 255, 255) in RGB color space, the altered pixel value i.e. (255 + $\Delta$, 255 + $\Delta$, 255 + $\Delta$) will be truncated back to the (255, 255, 255) leading to automatic destruction of watermark. Considering these challenges, the proposed method embeds the watermark inside the \textquotedblleft texture blocks\textquotedblright\ only using a content adaptive watermarking strength. 

The energy of each block is different. The watermark embedding strength will vary for each block. Generally, the watermark embedding strength $\alpha$ ranges from $0$ to $1$ and is dependent upon \emph{imperceptibility} and \emph{robustness}. Texture blocks with high as well as low energy will have little embedding capacity and hence, the watermark embedding strength should be less while texture blocks with moderate energy have comparatively high embedding strength and hence, the watermark embedding strength should be relatively high. In summary, watermark strength should be adaptive to the content. A few watermark embedding curve examples (continuous as well as staircase functions) are shown in Figure \ref{fig:waterStrength}. In the proposed method, we have used a staircase function. We embedded the watermark using DWT-based image watermarking algorithm at different $\alpha$ values for determining the watermark embedding strength for each category of block. After embedding a watermark, we manually inspected the watermarked blocks for imperceptibility while extracted the watermark using DWT-based image watermarking algorithm for checking the robustness. Results are summarized in Table \ref{table:alphaAdjust}. From Table \ref{table:alphaAdjust}, we can see that  $\alpha =0.1$ for CT blocks and $\alpha =0.2$ for PTPG as well as CG blocks lead to watermark detection while being imperceptible. Hence, we have used these values for embedding the watermark inside texture blocks. As non-texture blocks lead to watermark loss, we have set $\alpha=0$ for these blocks i.e. for CW and PT blocks. Further, we have tested these values of $\alpha$ with other image-based watermarking algorithms (see Table \ref{table:algoDetails}) and found that all the image watermarking algorithms are able to detect the watermark with above specified values of $\alpha$.
\begin{table}
\begin{minipage}{.5\textwidth}
\centering
\fbox{\includegraphics[width=0.9\columnwidth]{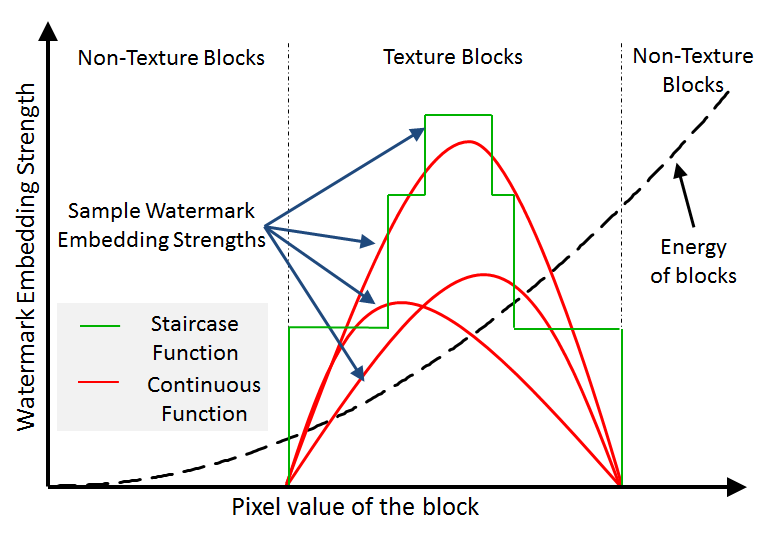}}
\captionof{figure}{Possible functions of watermark strengths (for the sake of illustration, we assume that block is homogenous i.e values of all pixels inside the block are same.)}
\label{fig:waterStrength}
\end{minipage}
\begin{minipage}{.49\textwidth}
\centering
\caption{Determination of $\alpha$}{
\begin{tabular}{|c|c|c|c|c|}
\hline
 & \multicolumn{2}{|c|}{Non-texture blocks} & \multicolumn{2}{|c|}{Texture blocks}\\
\cline{2-5}
$\alpha$ &	\textbf{CW}	& \textbf{PT}	& \textbf{CT} & \textbf{PTAG/CG} \\
\hline
0.5	& IND &	IND &	PD &	PD \\
\hline
0.25	& IND &	IND	& PD &	PD\\
\hline
0.2 &	IND &	IND &	PD & 	\textbf{ID}\\
\hline
0.1 &	IND &	IND &	\textbf{ID} & 	IND\\
\hline
0 &	\textbf{IND} &	\textbf{IND} &	IND	& IND\\
\hline
\multicolumn{5}{l}{IND - Imperceptible and not detectable}\\
\multicolumn{5}{l}{ID - Imperceptible and detectable}\\
\multicolumn{5}{l}{PD - Perceptible and detectable}\\
\multicolumn{5}{l}{$\alpha$ - Watermark Embedding Strength}\\
\end{tabular}}
\label{table:alphaAdjust}
\end{minipage}
\end{table}
%%%%%%%%%%%%%%%%%%%%%%%%%%%%%%%%%%%%%%%%%%%%%%%%%%%%%
\begin{table}[b!]
\centering
\caption{Notations used for representing image-based watermarking algorithms}{
\begin{tabular}{|l|c|}
\hline
\textbf{Watermarking Algorithm} & \textbf{Abbreviation}\\
\hline
DWT-based watermarking algorithm \cite{Reddy2005:eke} & Algo 1 \\
\hline
SVD-based watermarking algorithm \cite{RuizhenLiu2002:eke} & Algo 2 \\
\hline
DCT-based watermarking algorithm \cite{cox:eke} & Algo 3 \\
\hline
DWT-SVD based watermarking algorithm \cite{Ganic2004:eke} & Algo 4 \\
\hline
DCT-DWT-SVD based watermarking algorithm \cite{Navas2008:eke} & Algo 5\\
\hline
\end{tabular}
}
\label{table:algoDetails}
\end{table}
%%%%%%%%%%%%%%%%
\item \textbf{Watermark Extraction:}  To extract the watermark from suspected document files, the proposed method converts the document into pages. Let us say that $P'$ pages having dimension $N' \times M'$ exist in the suspected document file. The proposed method detects margins in $P'$ pages and generates cropped pages $P'_C$ using the method discussed in Section \ref{sssec:blockAnalysis}. The size of $P'_C$ pages might not be same as $P_C$ pages. As the proposed embedding method is block wise, the mismatch in the dimension of the original cropped page $P_C$ and the attacked cropped page $P'_C$ will lead to incorrect detection of blocks. This may result in loss of embedded information. To avoid such scenarios, the proposed method resizes $P'_C$ to the size of $P_C$ using bilinear interpolation \footnote{We kept the size of cropped page $P_C$ same across all the documents. We pass the size of $P_C$ as an argument to the extractor. In case, the size of $P_C$ and $P'_C$ do not match, we resize $P'_C$ to $P_C$ using bilinear interpolation.}. Once suspected file pages are resized, the proposed method segments $P'_C$ pages into blocks of dimension $n \times m$ and classify them into texture and non-texture blocks using the method discussed in Section \ref{sssec:blockAnalysis}. Now, the proposed method extracts the watermark from the texture blocks using the \textit{same watermarking method} which is used for embedding the watermark.
\end{itemize}
%%%%%%%%%%%%%%
\section{Evaluation Criteria}
\label{sec:criteria}
Unobtrusiveness and robustness are the two major requirements of any watermarking algorithm. The parameters used to assess the performance of the proposed method are discussed in this section.
\subsection{Quality Assessment Parameters}
\label{ssec:qualityAssessmentParameters}
The quality of the watermarked documents with respect to the original documents can be assessed using either subjective or objective assessment.
\begin{enumerate}
\itemsep 0em
\item {\bf Subjective Assessment:} In this type of assessment, readers are asked to respond to a questionnaire and rate their reading experience on a scale of 1 to 5 as shown in Table \ref{table:qualityScale}. Although this type of assessment provides the best result in comparison with objective assessment, it is time consuming and difficult to carry out due to cultural diversities and different perception abilities of individuals.
%=================================
\begin{table}[h!]
\centering
\caption{Quality Assessment Scale \protect\cite{Klaue2003:eke}}{
\begin{tabular}{|c|c|c|c|c|c|}
\hline
\textbf{Score} & 5 & 4 & 3 & 2 & 1 \\
\hline
\textbf{Quality} & Excellent & Good & Fair & Poor & Bad \\
\hline
\end{tabular}}
\label{table:qualityScale}
\end{table}
Let us assume that $s$ subjects participated in the survey and the rating of each subject is $r_{ij}$, where $j$ represents the number of subjects who participated in the survey of language $i$. A subjective Mean Opinion Score (MOS) can be computed as $MOS_i = \sum_{j=1}^{s}r_{ij}/s$.
\item {\bf Objective Assessment:} In this type of assessment, mathematical tools are used to estimate the quality of the watermarked content with respect to the original content. This type of assessment can be done easily as it does not have any human involvement. The proposed method uses Peak Signal to Noise Ratio (PSNR) and the Structural Similarity Index (SSIM) to measure the quality of the watermarked documents \cite{Chikkerur2011:eke}. Watermarked content with PSNR value greater than 30 dB \cite{Klaue2003:eke} or SSIM close to 1 \cite{Chikkerur2011:eke} is considered as indistinguishable from original content.\\
	The proposed method also uses another matrix, \textit{Normalized Correlation Coefficient} (NC) \cite{Lin2001:eke}, to measure the similarity between the original watermark $W$ and the extracted watermark $W_E$. The value of NC lies between 0 and 1 where 1 signifies that the original and extracted watermark are identical while 0 signifies that the original and extracted watermark are different. Since the proposed method embeds the watermark in each texture block, we shall be having as many extracted watermarks as number of texture blocks. The proposed method computes the overall correlation coefficient of the document using Algorithm \ref{algo:correlation}. For a given language, the correlation coefficient is an average of correlation coefficients of all the documents in that language.
\begin{algorithm}
\scriptsize
 \caption{Computing correlation coefficient for the entire document}
\textbf{Step 1:} Compute correlation coefficient for each block as:
$NC(i) = \frac{\sum \sum (W - \overline{W}) (W_E - \overline{W_E})}{\sqrt{\sum \sum(W - \overline{W})^2 \sum \sum(W_E - \overline{W_E})^2}}$, where $1\leq i \leq b$ and $\overline{(.)}$ represents a mean of matrix element.
\BlankLine
\textbf{Step 2:} Now, we sort NC in descending order as:
$NC_{sort} = sortDescending(NC)$
\BlankLine
\textbf{Step 3:} Now, we compute the overall correlation coefficient of a document by averaging first 25\% values from $NC_{sort}$ as:
$NC_{overall} = average(NC_{sort}(j)), 1 \leq j \leq 0.25b$
\BlankLine
\label{algo:correlation}
\end{algorithm}
\end{enumerate}
%%%%%%%%%%%%%%%%%%%%%%%%%%%5
\subsection{Robustness Assessment Parameters}
\label{ssec:robustnessAssessment}
The following attacks have been conducted to test the performance of the proposed method:
\begin{itemize}
\item \textit{Print Screen:} Duplicate copies of digital documents can be easily created using \textquotedblleft print screen\textquotedblright\  option. Print screen is a \textquotedblleft system property\textquotedblright\ and requires expensive DRM solutions, such as CopySafe PDF \cite{CopySafe2014:eke}, to disable it. In the absence of DRM solutions, print screen attack can be conducted in following ways:
\begin{enumerate}
        \item \textit{Using a screen shot:} This is the easiest way of capturing the content which is currently being displayed on the screen. A malicious user can press the \textquotedblleft Take a screen shot\textquotedblright\ option in the document file and manually select the page. This will generate an image which can be converted easily to a PDF file. These PDF files can be distributed or shared illegally. Manual print screen leads to \textit{scaling} as well as \textit{cropping} attacks.
        \item \textit{Using software:} Software and Application Programming Interfaces, such as ClearestPDFToImage \cite{Clearest2014:eke}, allow a user to easily manipulate the PDF documents. PDF files at different quality levels can be generated using these softwares and APIs. To conduct this attack, we have used ClearestPDFToImage software \cite{Clearest2014:eke}. Further, we have considered 3 quality levels for documents: (i) Low, (ii) Normal, and (iii) Good. Print screen using software leads to \textit{scaling} attack, \textit{compression} attack, and slight rotations in the text. Down scaling occurs when the attack is conducted at low quality while up scaling occurs when the attack is conducted at normal as well as good quality.
    \end{enumerate}
We have assumed that the PDF generated through print screen attack will not be useful if the text content (text, graphs, images, tables, etc.) is cropped. Hence, we have conducted the print screen experiment while preserving maximum of the text content.
%========================
\item \textit{Print and Scan Attack:} The print and scan process has been widely studied in literature for the robustness assessment of text watermarking methods. Printing is a \textquotedblleft document property\textquotedblright\ which can be disabled easily using PDF processing tools such as \textit{JPedal} \cite{JPedal2014:eke}. Though the printing option can be easily disabled using PDF processing tools, there are other ways to enable the printing option in a PDF file. For instance, a malicious user can try to enable the printing option by transcoding a PDF file (PDF to PS to PDF). In such a case, the malicious user will be able to ``print and scan" the document. To study the robustness of the proposed method against ``print and scan" process, we printed the documents and scanned them at different DPI values i.e. 100, 200 and 300. In our experiments, we have used Canon's black and white printer cum scanner (ImageRunner 2520).
\item \textit{Text alteration:} Malicious users can try to remove the watermark by altering the content of the text document. To check the robustness of the proposed method, we modified the content of the text document by: (i) transforming the
sentence, (ii) replacing words with synonyms, (iii) highlighting the text, and (iv) manually strike through (multiple times) the word and then writing it back on top of strike through word.
%========================================================
\item \textit{Stitching Attack:} In the proposed method, we treat each page of the document as an image. A malicious user may try to crop the pages and stitch them back to produce an exact or similar copy. Stitching of pages may lead to automatic destruction of the embedded watermark. To study the robustness of the proposed method, we have conducted this attack in two ways:
\begin{enumerate}
\item \textit{Column/Row-wise stitching:} To conduct this attack, we cropped the pages either column or row-wise and then stitched  back the cropped portions to produce nearly an exact copy. Row/Column-wise stitching leads to multiple changes in a page such as reduced space between columns, lines may be shifted up/down, and reduced space between paragraphs. Figure \ref{fig:colstich} shows a snapshot of the small portion of the page obtained by column-wise stitching. We can clearly see that column-wise stitching can lead to a change in space between columns as well shifting of lines.
 Such changes can lead to automatic destruction of the watermark.
\begin{figure}[t!]
\centering
\subfigure[Before]{\includegraphics[width=2.75cm]{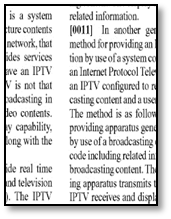}}
\hspace{0.5cm}
\subfigure[After]{\includegraphics[width=2.75cm]{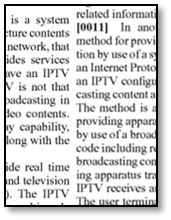}}
\caption{Example of column-wise stitching}
\label{fig:colstich}
\end{figure}
%================ 
\item \textit{Page-wise stitching:} To conduct this attack, we stitched two pages to form one page, as shown in Figure \ref{fig:pagestich}. Page-wise stitching might not lead to a seamless reading experience. When two pages are stitched to form one page, the original pages are resized to almost half of their original width resulting in a degradation in quality of the document.
\begin{figure}[t!]
\centering
\subfigure[Page 1]{\makebox[4cm]{\includegraphics[width=2.75cm]{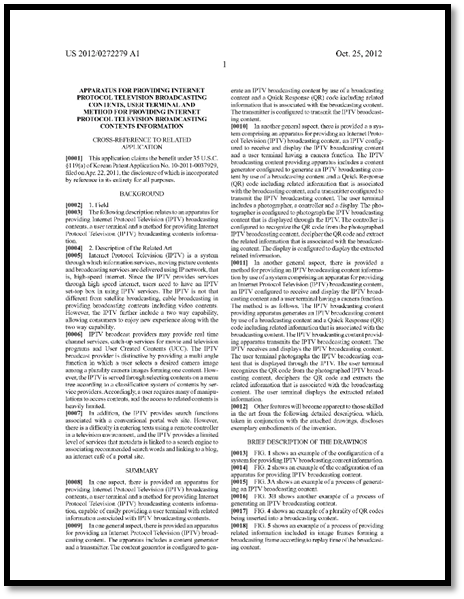}}}
\hspace{0.5cm}
\subfigure[Page 2]{\makebox[4cm]{\includegraphics[width=2.75cm]{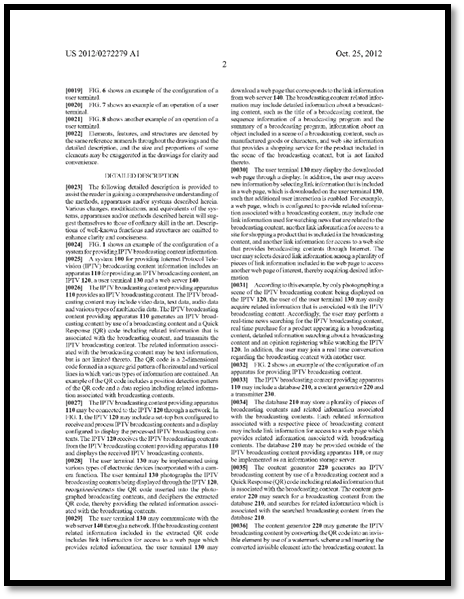}}}
\hspace{0.5cm}
\subfigure[After Stitching]{\makebox[4cm]{\includegraphics[width=2.75cm]{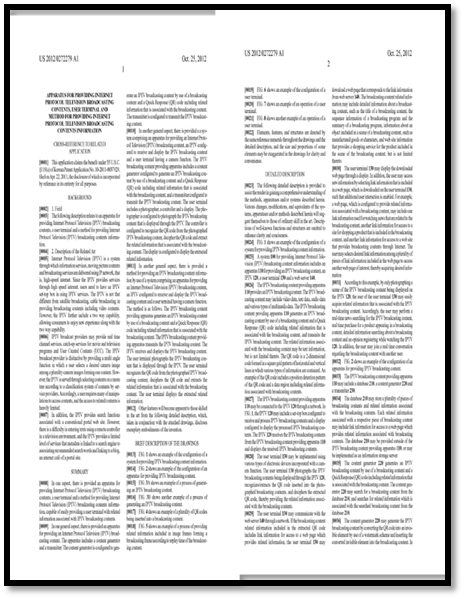}}}
\caption{Example of page-wise stitching: (a-b) Two consecutive pages of a document (c) Resultant page obtained after stitching (a) and (b)}
\label{fig:pagestich}
\end{figure}
\end{enumerate}
%===========================================
%========================
\item \textit{Signal Processing Attacks:}  As the proposed method treats each document as a set of images, image-based attacks are possible on the text document. To check the robustness of the proposed method under such attacks, we have conducted following attacks:
\begin{enumerate}
\item {\it Skew or Rotation:} Skew is inevitably introduced while scanning the document. Under such circumstances, watermarking algorithms might fail to detect the watermark. To study the robustness of the proposed method under this attack, we varied the rotation angle from $0.1^\circ$ to $10^\circ$. As we will see in Section \ref{sec:results}, the proposed method fails to extract the watermark in presence of skew. Since skew detection and correction is a well researched area and several robust algorithms exist in literature for detecting and correcting the skew, we integrated a well known skew detection and correction algorithm proposed by \cite{Cao2003:eke} in our framework. Results of skew detection are given in Figure \ref{fig:skewCorrection}. We can clearly see that the method is able to detect the skew with an error of $\pm 0.03^\circ$.

\begin{figure}
\centering
\subfigure[]{\fbox{\includegraphics[width=4.25cm]{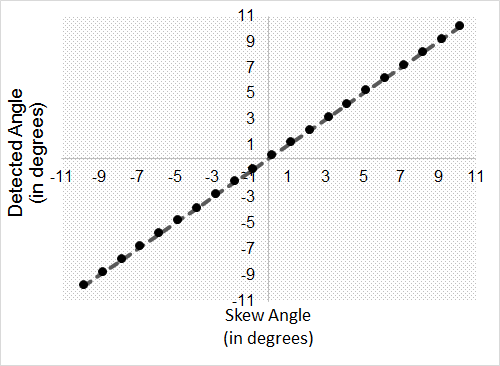}}}%
\hspace{1cm}
\subfigure[]{\fbox{\includegraphics[width=4.25cm]{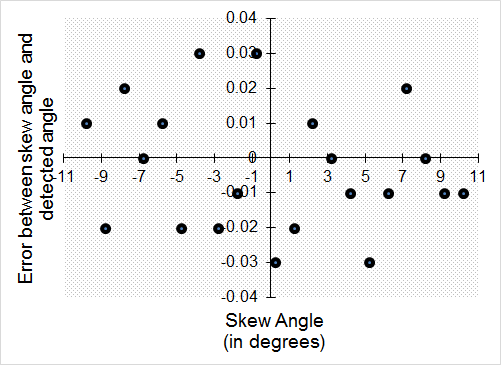}}}%
\caption{Results of skew detection: (a) Detected angle and (b) Error between detected angle and actual angle of rotation}
\label{fig:skewCorrection}
\end{figure}

\item {\it Compression:} Compression is a technique to reduce the storage space required by any media. As the proposed method treats the pages of a text document as images, this attack becomes critical for analysis. Compression may lead to a drastic reduction in storage space while introducing subtle variations in the document. To conduct this attack, we have varied the JPEG quality factor from $90$ to $10$ where $90$ denotes very low compression while $10$ denotes high compression.
        \item {\it Noise:} Generally, the watermark is considered as a noise to the signal in which it is embedded. Addition of extra noise (apart from the watermark) to the content may lead to destruction of the watermark. To check the robustness of the proposed method against such scenarios, we have added Gaussian noise $G$ to each page of the watermarked text documents $P_W$ to create the noisy page $P_N$. Noise variance $\sigma$ is varied from $0.1$ to $10$.
\end{enumerate}
\end{itemize}
%%%%%%%%%%%%%%%%%%%%%%%%%%%%%
\section{Experimental Results}
\label{sec:results}
%%%%%%%%%%%%%%%%%%%%%%%%
\subsection{Experimental Environment}
\label{ssec:expEnvironment}
\begin{figure}[t!]
\centering
\subfigure[No. of Documents per Language]{\fbox{\includegraphics[width=5cm]{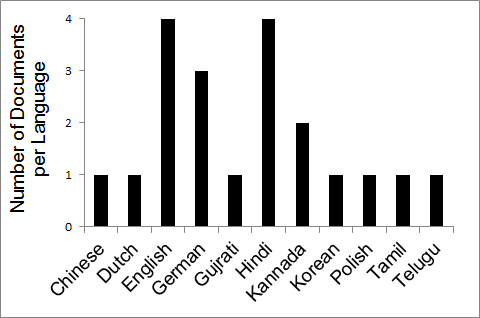}}}
\hspace{1cm}
\subfigure[No. of Pages per Language]{\fbox{\includegraphics[width=5cm]{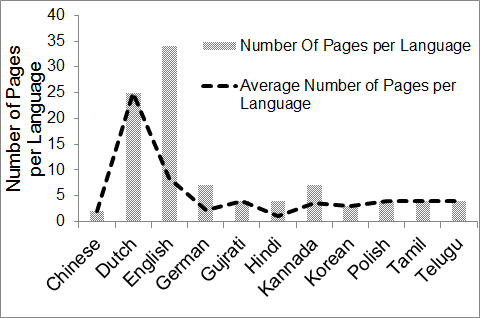}}}
\caption{Details of text document files}
\label{fig:textDocDetails}
\end{figure}

To test the performance of the proposed method, $20$ text Portable Document Files (PDF's) are used. These files differ in terms of language (English, Hindi, Tamil, Chinese, etc.), number of pages ($1$ to $35$), content type (text, graphs, tables, images, equations, color background, etc.), and page size. The details of the text documents are given in Figure \ref{fig:textDocDetails}. We have conducted the experiments on a machine having AMD Athlon II $\times$2 245 2.90 GHz processor and 4 GB RAM.
\subsection{Setting of Block Size}
\label{ssec:settingBlockSize}
The proposed method divides the text document into blocks. The size of the block has a direct impact on the capacity of the watermarking algorithm as well as the amount of information in the texture blocks.
\begin{itemize}
\item \textit{Impact of block size on Texture Information: }Texture blocks of different sizes are shown in Figure \ref{fig:textureBlocksDifferent}. From Figure \ref{fig:textureBlocksDifferent}, it is clear that a block of size $64 \times 64$ is able to carry much less texture information while other block sizes are able to carry a significant amount of texture information.
\begin{figure}[t!]
\centering
\subfigure[]{\fbox{\label{fig:512}\includegraphics[scale=0.15]{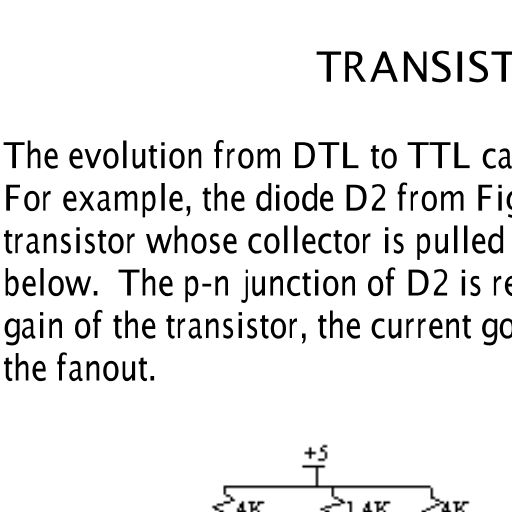}}}
\hspace{0.5cm}
\subfigure[]{\fbox{\label{fig:256}\includegraphics[scale=0.15]{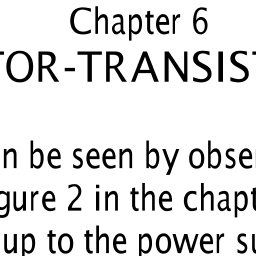}}}
\hspace{0.5cm}
\subfigure[]{\fbox{\label{fig:128}\includegraphics[scale=0.15]{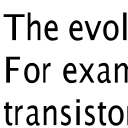}}}
\hspace{0.5cm}
\subfigure[]{\fbox{\label{fig:64}\includegraphics[scale=0.15]{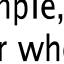}}}
\caption{Texture blocks of different sizes: (a) 512 $\times$ 512, (b) 256 $\times$ 256, (c) 128 $\times$ 128, (d) 64 $\times$ 64}
\label{fig:textureBlocksDifferent}
\end{figure}
%================================================
\item \textit{Impact of block size on Capacity: }Capacity of any algorithm represents the amount of information it can carry reliably.\\
\textbf{Algo 1:} Capacity of Algo1 is dependent upon the DWT decomposition level. Table \ref{table:capacity} provides the details of capacity of Algo1. The order of the capacity of Algo1 is:
    \begin{equation}
    Level-1 > Level-2 > Level-3
    \end{equation}
    We embedded the watermark in documents using Algo 1 at 3 different levels. On manual inspection of the text documents, we found that the watermark is visible in Level-1 and Level-2 but not in Level-3. Hence, we selected Level-3 as the decomposition level for Algo 1. Further, Algo 1 requires $\mathcal{O}(XY)$ operations for a block having dimension $X \times Y$ \cite{golub1996:eke}. Although a block size of 512 $\times$ 512 has more information carrying capacity then other block sizes in Table \ref{table:capacity}, it has highest computational complexity. Block sizes of 64 $\times$ 64 and 128 $\times$ 128 have much less watermark carrying capacity while block size of 256 $\times$ 256 is able to carry a significant amount of information. Generally, the watermarking algorithm should have a suitable tradeoff between imperceptibility, watermark carrying capacity, and computational complexity. Considering these parameters, we selected 256 $\times$ 256 as the block size for Algo 1.
\begin{table}[t!]
\begin{minipage}{0.5\textwidth}
\centering
\caption{Capacity of Algo 1}{
\begin{tabular}{|c|c|c|c|}
\hline
\multirow{2}{*}{\textbf{Block size}} & \multicolumn{3}{|c|}{\textbf{Algo1}}\\
\cline{2-4}
& Level - 1 & Level - 2 & Level - 3 \\
\hline
512 $\times$ 512 & 256 $\times$ 256 & 128 $\times$ 128 & 64 $\times$ 64\\
\hline
256 $\times$ 256 & 128 $\times$ 128 & 64 $\times$ 64 & 32 $\times$ 32\\
\hline
128 $\times$ 128 & 64 $\times$ 64 & 32 $\times$ 32 & 16 $\times$ 16\\
\hline
64 $\times$ 64 & 32 $\times$ 32 & 16 $\times$ 16 & 8 $\times$ 8\\
\hline
\end{tabular}}
\label{table:capacity}
\end{minipage}
\begin{minipage}{0.5\textwidth}
\centering
\fbox{\includegraphics[scale=1.25]{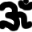}}
\captionof{figure}{Original Watermark ($32 \times 32$)}
\label{fig:watermark}
\end{minipage}
\end{table}

\textbf{Algo 2:} For Algo 2, the capacity is directly proportional to the block size. This is because the singular values of the watermark are embedded inside the singular values of the blocks. As the watermark is binary in nature, its singular values are small in comparison to singular values of the block and in most cases, the last 6 to 7 singular values are equal to zero. Furthermore, Algo 2 requires $\mathcal{O}(XY^2 + X^2Y)$ operations for a block size of dimension $X \times Y$ \cite{golub1996:eke}. Considering the computational complexity, watermark carrying capacity, and the amount of information in the texture blocks, we selected 128 $\times$ 128 as block size for Algo 2.

\textbf{Algo 3:} For global DCT-based watermarking algorithm, the capacity is directly proportional to the block size\footnote{We have not used local DCT-based watermarking algorithm in which image is divided into a size of $8\ \times\ 8$ and watermark is embedded inside these $8\ \times\ 8$ blocks. The reason is quite obvious. As we saw in Figure \ref{fig:textureBlocksDifferent}, the information carrying capacity reduces with the decrease in block-size. If we reduce the block-size to $8\ \times\ 8$, then majority of the blocks are either black or white. If we try to embed the watermark in such blocks, then watermark is either automatically destructed or becomes visible.}. Also, Algo 3 requires $\mathcal{O}(XY)$ operations for a block size of dimension $X \times Y$. Considering the computational complexity, watermark carrying capacity, and the amount of information in the texture blocks, we selected 128 $\times$ 128 as block size for Algo 3.

\textbf{Algo 4 \& Algo 5:} Algo 4 \& Algo 5 are combinations of Algo 1, Algo 2, and Algo 3. Capacity of Algo 4 and Algo 5 is equal to $\frac{1}{4}^{th}$ of the block size. However, computational complexities of Algo 4 and Algo 5 are different. On a block of dimension $X \times Y$, Algo 4 require $\mathcal{O}(XY)$ and $\mathcal{O}(XY^2 + X^2Y)$ operations for computing DWT and SVD respectively while Algo 5 require $\mathcal{O}(XY)$, $\mathcal{O}(XY)$, and $\mathcal{O}(XY^2 + X^2Y)$ operations for computing DWT, DCT, and SVD respectively. Considering computational complexity, embedding capacity, and amount of information in the texture blocks; we selected 128 $\times$ 128 as block size for Algo 4 and Algo 5.

For a comparative study of different image watermarking algorithms, the proposed method used a watermark of dimension $32 \times 32$, shown in Figure \ref{fig:watermark}.
\end{itemize}
%==========================================
\subsection{Impact on File Size}
\label{ssec:fileSize}
The watermark acts like a noise to the document and embedding of watermark will have an impact on the document size. As the proposed method is block-wise approach, we compared the size of the blocks before and after watermarking. We found that the proposed method increases the size of CT blocks by approximately 0.7\% while the size of PTPG and CG blocks by approximately 1\%. The increase in block size is quite less and hence, we can say that the proposed method is practical.

%============================================
\subsection{Quality Assessment}
\label{ssec:qualityAssessment}
Embedding of a watermark inside the text document may degrade the quality of the text document. We have used the parameters discussed in Section \ref{ssec:qualityAssessmentParameters} to assess the quality of the watermarked text documents with respect to the original text documents.
%%%%%%%%%%%%%%%%%%%%%%
\subsubsection{Impact of Watermarking on Different Elements of Text Document}
\label{sssec:impact}
\begin{figure*}[b!]
\centering
\includegraphics[width=0.8\columnwidth]{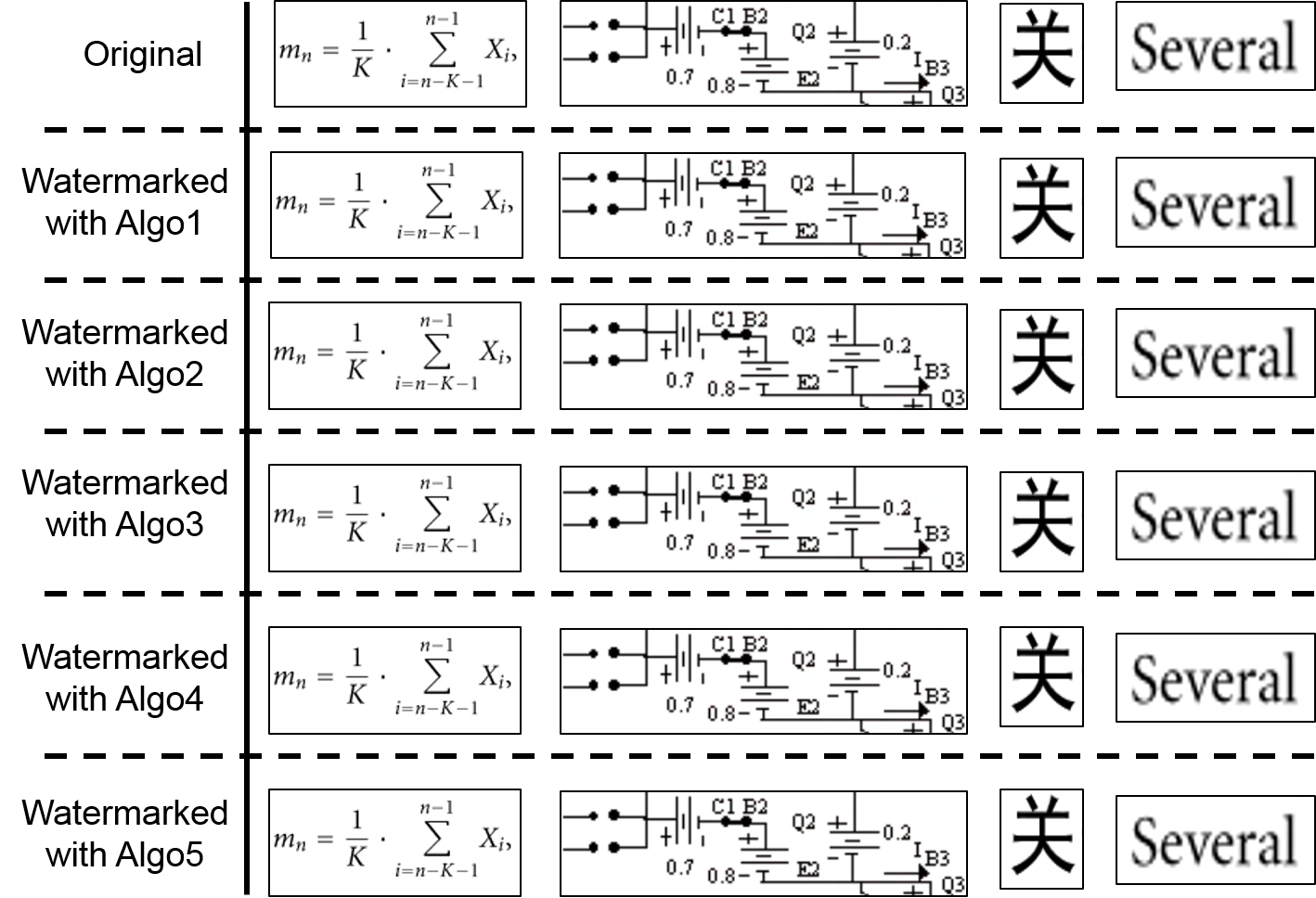}
\caption{Impact of Watermarking on different elements (characters, equations, etc.) of a text document}
\label{fig:effectCharWatermarking}
\end{figure*}

A text document contain different elements such as text, equations, images, graphs, and charts. Figure \ref{fig:effectCharWatermarking} shows the snapshots of a few elements present in the text document. From Figure \ref{fig:effectCharWatermarking}, it is clear that watermarking of the text document with different watermarking algorithms does not deteriorate the quality of different elements present inside the text documents.
%======subjective assessment=========
%%%%%%%%%%%%%%%%%%%%%%%%
%======================================================
%%%%%%%%%%%%%%%%%%%%%%%%%%%%%
\subsubsection{Subjective Assessment}
\label{sssec:subjectiveAssessment}
Subjective assessment relates to the actual reading experience. Carrying out such reading assessments requires sufficient cultural, racial, and gender diversity of readers. Hence, we have limited our reading assessment as an informal one, limited to  \textbf{24} subjects. Subjects are under-graduate students, graduate students, researchers, faculty, and employees. Among these $24$ subjects, a few of them participated in multiple languages: $10$ subjects participated in $2$ languages, $2$ subjects participated in $3$ languages, and $1$ subject participated in 4 languages. 

\begin{figure*}[b!]
\centering
\subfigure[Subjective Assessment]{\fbox{\label{fig:mosSubjectiveAssesssment}\includegraphics[width=4.25cm]{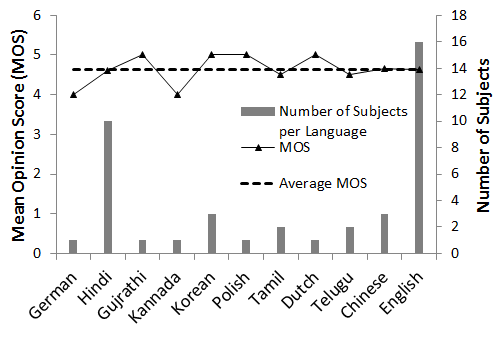}}}
\hspace{0.001cm}
\subfigure[PSNR]{\fbox{\label{fig:objPSNR}\includegraphics[width=4.25cm]{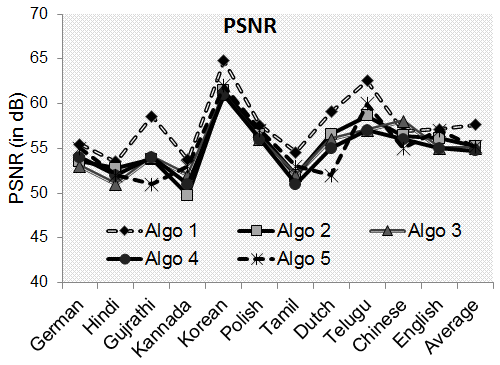}}}%
\hspace{0.001cm}
\subfigure[SSIM]{\fbox{\label{fig:objSSIM}\includegraphics[width=4.25cm]{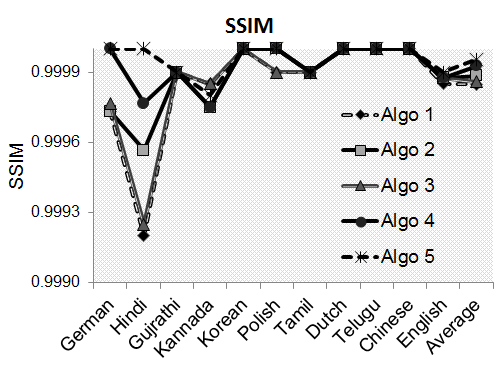}}}%
\caption{Quality Assessment: (a) Subjective Assessment, (b-c) Objective Assessment}
\label{fig:qualityAssessment}
\end{figure*}

We shared the original as well as the watermarked documents (watermarked using different algorithms) with the subjects and asked the following questions:
\begin{enumerate}
\itemsep 0em
\item Are there any distortions in the characters (such as discontinuities, missing links in characters, etc.) due to watermarking?
\item Is there any loss of information in graphs, images, and tables due to watermarking?
\item Are there any other visual distortions in the document due to watermarking?
\item Rate your overall reading experience.
\end{enumerate}
As per the survey, subjects were not able to identify any difference between the original and watermarked documents. Based on the ratings of subjects, the average MOS for each language is computed and is shown in Figure \ref{fig:mosSubjectiveAssesssment}. Average MOS across all the languages is $4.63$. Average MOS values clearly indicate that watermarked documents obtained using different watermarking algorithms are of good quality and the subjects were not able to distinguish between the original and watermarked documents.
\subsubsection{Objective Assessment}
\label{sssec:objectiveAssessmentResults}
Figures \ref{fig:objPSNR} and \ref{fig:objSSIM} contain the PSNR as well as SSIM index values. From Figures \ref{fig:objPSNR} and \ref{fig:objSSIM}, it is clear that PSNR values for all watermarking algorithms under consideration are greater than $45$ dB while the SSIM index values are greater than $0.99$ (close to $1$). These values clearly indicate that the degradations introduced to the original document due to watermarking are minimal. This is consistent with the observations obtained from the subjective assessment.
% degradations resulted in original document due to watermarking are very less.
%%%%%%%%%%%%%%%%%%%%%%%%%%%%%
\subsection{Robustness Assessment}
\label{ssec:robustResults}
The robustness of the proposed method under different attacks is discussed below:
\subsubsection{Without Any Attack}
\label{sssec:withoutAttack}
\begin{figure}[b!]
\begin{minipage}{.35\textwidth}
\centering
\fbox{\includegraphics[width=4cm]{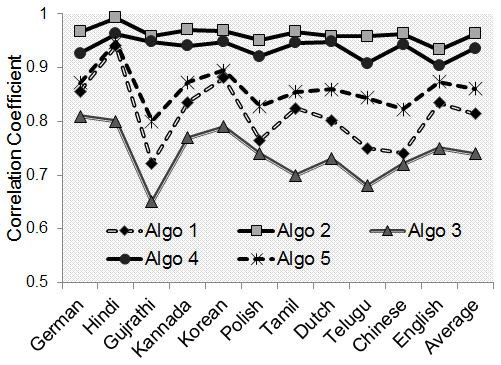}}
\caption{Robustness without any Attack}
\label{fig:noAttack}
\end{minipage}
\hspace{0.5cm}
\begin{minipage}{.65\textwidth}
\centering
\fbox{{\includegraphics[width=4cm]{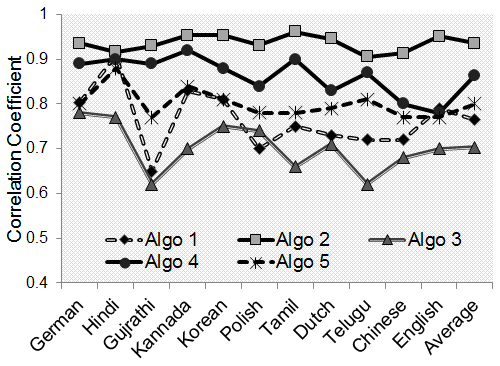}}}
\caption{Robustness against print screen attack conducted using Screen Shot}
\label{fig:scrnShotAttack}
\end{minipage}
\end{figure}

Figure \ref{fig:noAttack} shows the performance of watermarking algorithms without any attack. For all languages under consideration, the average value of correlation coefficient across different languages for Algo 1, Algo 2, Algo 3, Algo 4,and Algo 5 are 0.81, 0.96, 0.74, 0.94, and 0.86 respectively.
%%%%%%%%%%%%%%%%%%
\subsubsection{Print Screen Attack}
\label{sssec:printScreenAttack}
\begin{figure*}[t!]
\centering
\subfigure[Low]{\fbox{\includegraphics[width=4.25cm]{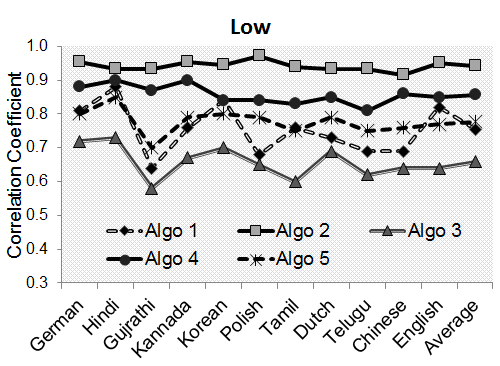}}}%
\hspace{0.01cm}
\subfigure[Normal]{\fbox{\includegraphics[width=4.25cm]{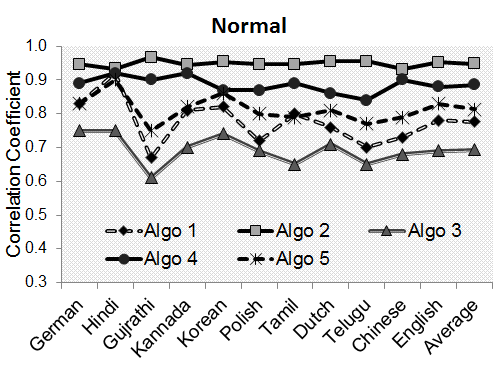}}}%
\hspace{0.01cm}
\subfigure[Good]{\fbox{\includegraphics[width=4.25cm]{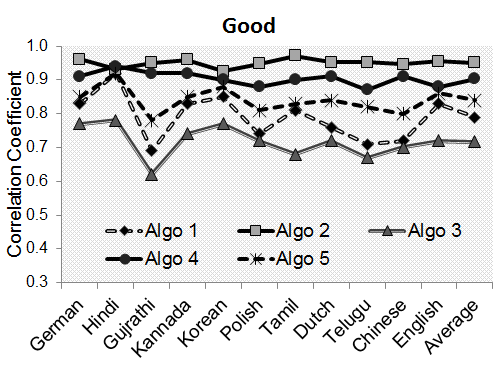}}}
\caption{Robustness against Print Screen Attack conducted using software}
\label{fig:prscrSoft}
\end{figure*}

Print screen attack can either scale or crop or slightly rotate the content. Manual print screen attack leads to scaling as well as cropping while print screen attack using software leads to scaling, compression, and slight rotations.
\begin{itemize}
\item \textit{Using screen shot option\footnote{While conducting this attack, we tried to preserve around 95\% of the text content.}:} Figure \ref{fig:scrnShotAttack} shows the performance of different algorithms under print screen attack conducted using the screen shot option. For all languages under consideration, the average value of correlation coefficient across different languages for Algo 1, Algo 2, Algo 3, Algo 4,and Algo 5 are 0.76, 0.94, 0.70, 0.86, and 0.80 respectively.
\item \textit{Using Software:} Figure \ref{fig:prscrSoft} shows the correlation coefficient of the extracted watermark with respect to the embedded watermark.  It is clear from Figure \ref{fig:prscrSoft} that the performance of watermarking algorithm remains almost constant with change in quality of the document from low to good. The average correlation coefficient value across all languages for Algo 1, Algo 2, Algo 3, Algo 4, and Algo 5 are [0.75, 0.79], [0.94, 0.95], [0.66, 0.72], [0.86, 0.90], and [0.78, 0.84] respectively when the quality of the document is changed from low to good.
\end{itemize}
%============
\subsubsection{Print and Scan Attack}
\label{sssec:printScanAttack}
\begin{figure*}[t!]
\centering
\subfigure[DPI - 100]{\label{fig:prScanDWT}\fbox{\includegraphics[width=4.25cm]{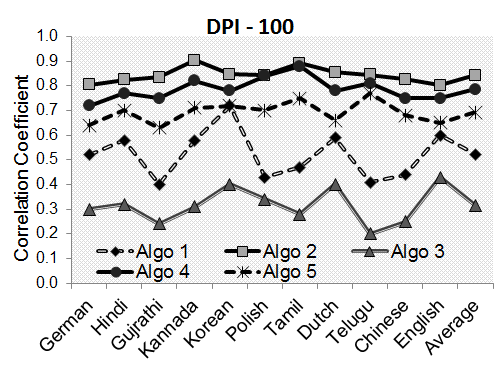}}}%
\hspace{0.01cm}
\subfigure[DPI - 200]{\label{fig:prScanSVD}\fbox{\includegraphics[width=4.25cm]{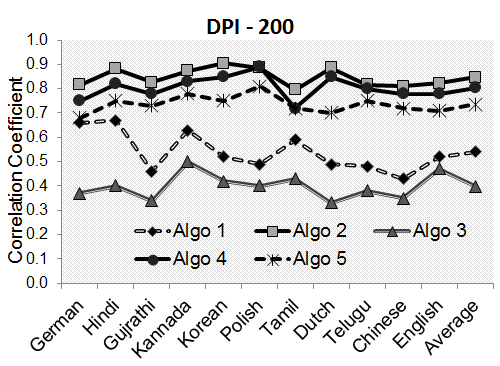}}}%
\hspace{0.01cm}
\subfigure[DPI - 300]{\label{fig:prScanOverall}\fbox{\includegraphics[width=4.25cm]{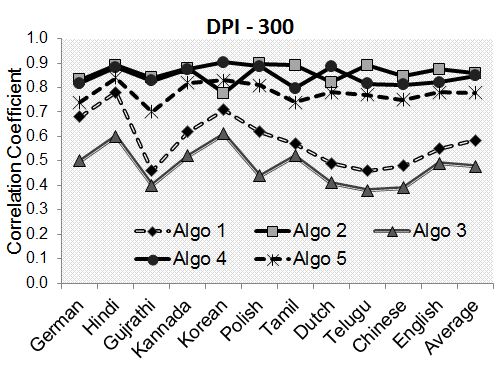}}}
\caption{Robustness against Print and Scan Attack at different DPI values}
\label{fig:printScan}
\end{figure*}

Figure \ref{fig:printScan} shows the performance of different watermarking algorithms under ``print and scan" attack at different DPI values. For all the documents under study, the average value of correlation coefficient across different languages for Algo 1, Algo 2, Algo 3, Algo 4, and Algo 5 are [0.52, 0.58], [0.84, 0.86], [0.32, 0.48], [0.79, 0.85], and [0.69, 0.78] respectively. The variation in correlation coefficient is because of the variation in content. We observed that the correlation coefficient values are lower for documents in which most of the content is colored graphics. There is a significant loss of information when we printed colored documents using a black and white printer which resulted in lower correlation coefficient of the extracted watermark.
%============
\subsubsection{Text Alteration}
\label{sssec:textModify}
\begin{figure}[t!]
\centering
\subfigure[]{\label{fig:textExample}\fbox{\includegraphics[width=4.75cm]{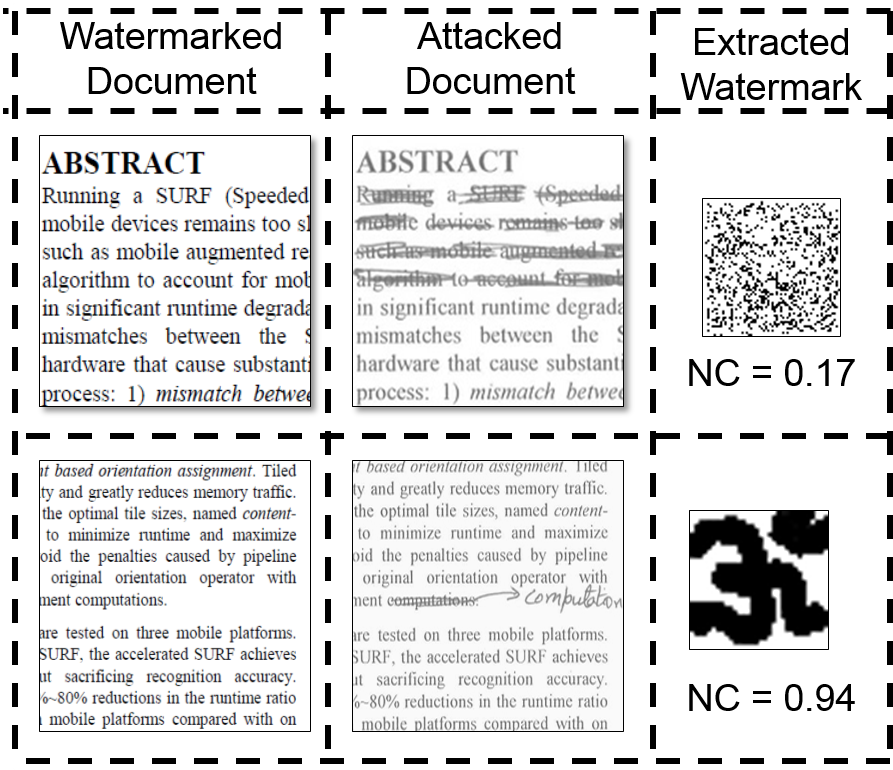}}}
\vspace{0.001cm}
\subfigure[]{\label{fig:textMod}\fbox{\includegraphics[width=4cm]{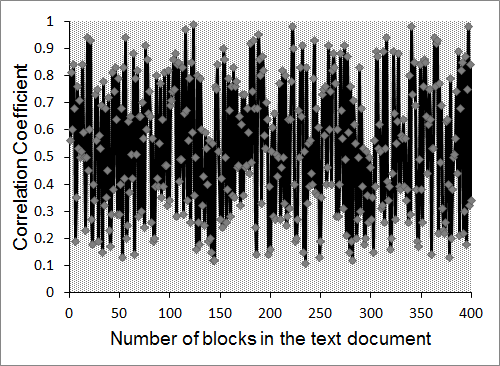}}}
\vspace{0.001cm}
\subfigure[]{\label{fig:textAlterFull}\fbox{\includegraphics[width=4cm]{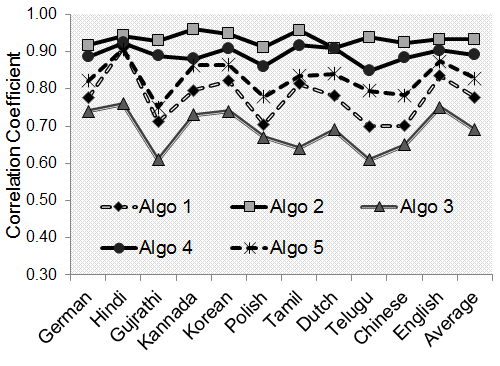}}}
\caption{Results of text modification: (a) Snapshots of attacked block and extracted watermark (using Algo 2), (b) Correlation coefficient of extracted watermarks from all blocks of a document (using Algo 2) and (c) Correlation coefficient of watermarks extracted using different watermarking algorithms across different languages}
\end{figure}

Figure \ref{fig:textMod} results of text document on one of the English document. The document contains 400 texture blocks in which watermark was embedded. We modified the content of the text document by: (i) transforming the sentence, (ii) replacing words with synonyms, (iii) highlighting the text, and (iv) manually strike through (multiple times) the word and sentences. Snapshots of two attacked blocks and extracted watermarks from these blocks are shown in Figure \ref{fig:textExample}. We noticed that the proposed method is able to detect the watermark from the blocks with minor alterations (such as strike though a single word or synonym substitution) while fails to extract the watermark from blocks with major alterations (e.g. strike though a sentence or sentence transformation).

We conducted the similar experiments on different documents and results are summarized in Figure \ref{fig:textAlterFull}. For all the documents under study, the average value of correlation coefficient across different languages for Algo 1, Algo 2, Algo 3, Algo 4, and Algo 5 are 0.78, 0.93, 0.69, 0.89, and 0.83 respectively. These values clearly indicate that the proposed method is able to sustain this attack.
\begin{figure*}[t!]
\centering
\subfigure[Column/Row-wise Stitching]{\makebox[5cm]{\fbox{\includegraphics[width=4.25cm]{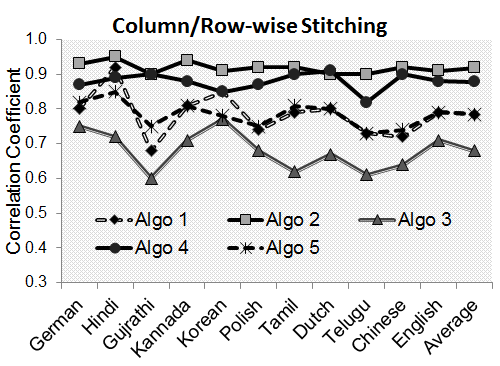}}}}
\hspace{1cm}
\subfigure[Page-wise Stitching]{\fbox{\includegraphics[width=4.25cm]{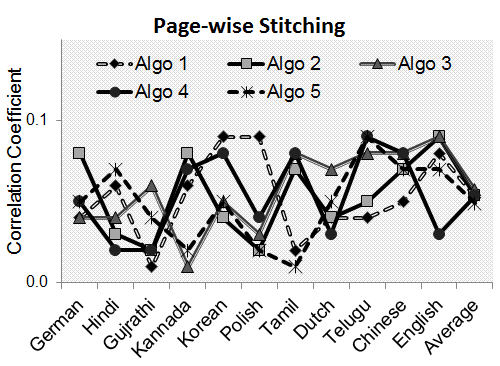}}}%
\caption{Robustness against Stitching Attack}
\label{fig:stitchAttack}
\end{figure*}

\subsubsection{Stitching Attack}
\label{sssec:stichAttack}
Figure \ref{fig:stitchAttack} shows the performance of different algorithms under ``stitching" attack. For all documents under consideration, the average value of correlation coefficient across different languages for Algo 1, Algo 2, Algo 3, Algo 4, and Algo 5 are 0.78, 0.92, 0.68, 0.88, and 0.78 respectively. These values indicate that the algorithms are able to handle the row/column-wise stitching attack. On the other hand, the proposed method is not able to handle the page-wise stitching attack (correlation coefficient values of the extracted watermarks using different watermarking algorithms are close to zero). Due to page-wise stitching, the content inside the blocks changed and hence, the proposed method failed to extract the watermark. It is worth noting that the proposed method is able to classify the blocks into different categories such as CW, PT, and CG under stitching attack (row/column-wise as well as page-wise stitching).
%==============================================================
\subsubsection{Rotation attack}
\label{sssec:rotationAttack}
\begin{figure}[t!]
\centering
\subfigure[Rotation = 1 $^{\circ}$]{\fbox{\includegraphics[width=4.25cm]{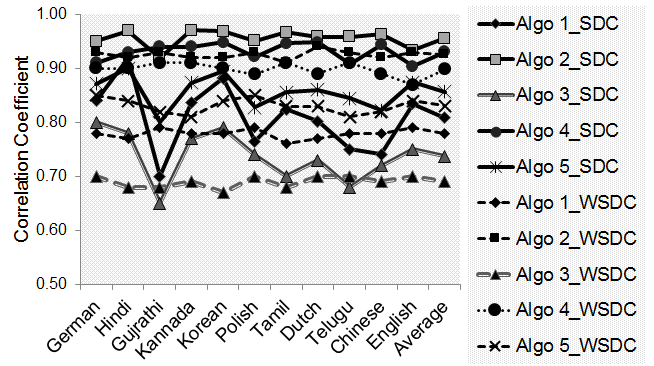}}}%
\vspace{0.001cm}
\subfigure[Rotation = 5 $^{\circ}$]{\fbox{\includegraphics[width=4.25cm]{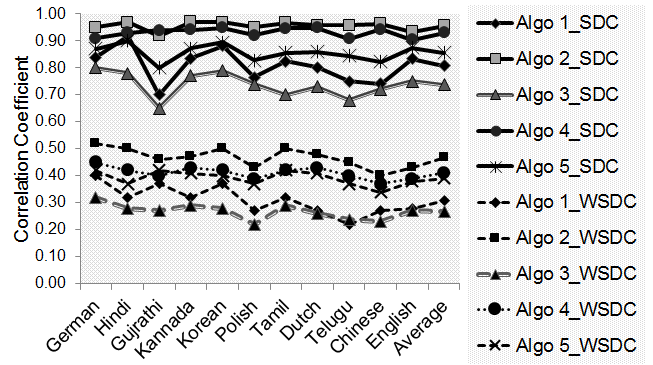}}}%
\vspace{0.001cm}
\subfigure[Rotation = 10 $^{\circ}$]{\label{fig:rotationEffectOverall}\fbox{\includegraphics[width=4.25cm]{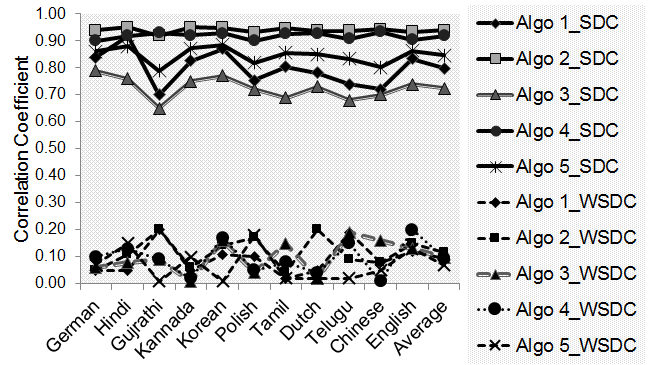}}}%
\caption{Robustness against Rotation Attack (SDC -with skew detection and correction; WSDC - Without skew detection and correction)}
\label{fig:rotationAttackCorrelation}
\end{figure}

Figure \ref{fig:rotationAttackCorrelation} shows the correlation coefficient of the extracted watermark with respect to the embedded watermark at different rotation angles. As we increase the angle of rotation, the method fails to detect the watermark in absence of skew detection and correction algorithm. However, the robustness of the proposed method increases after skew detection and correction. For all documents under consideration, the average value of correlation coefficient (after skew detection and correction) across different languages for Algo 1, Algo 2, Algo 3, Algo 4, and Algo 5 are between [0.80, 0.81], [0.94, 0.96], [0.73, 0.74], [0.92, 0.93], and [0.85, 0.86] respectively when we varied the degree of rotation from $1^\circ$ to $10^\circ$. 

\subsubsection{Compression Attack}
\label{sssec:compressAttack}
\begin{figure}[t!]
\centering
\subfigure[Q-factor = 90]{\makebox[3cm]{\fbox{\includegraphics[width=1.75cm]{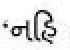}}}}%
\hspace{1cm}
\subfigure[Q-factor = 50]{\makebox[3cm]{\fbox{\includegraphics[width=1.75cm]{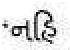}}}}%
\hspace{1cm}
\subfigure[Q-factor = 10]{\makebox[3cm]{\fbox{\includegraphics[width=1.75cm]{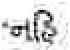}}}}%
\caption[Effect of compression on words and characters]{Effect of compression on words and characters (this word is from the Gujrati Language)}
\label{fig:compEffectChar}
\end{figure}

\begin{figure}[t!]
\centering
\subfigure[Q-10]{\fbox{\includegraphics[width=4.25cm]{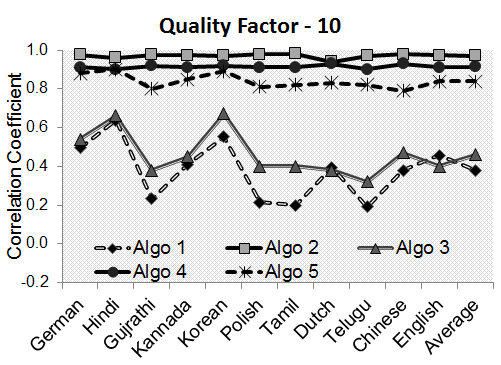}}}%
\vspace{0.001cm}
\subfigure[Q-50]{\label{fig:compEffectSVD}\fbox{\includegraphics[width=4.25cm]{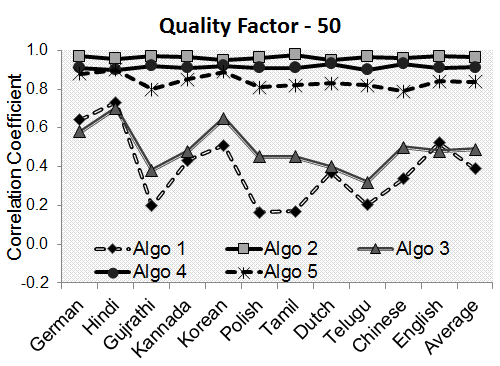}}}%
\vspace{0.001cm}
\subfigure[Q-90]{\label{fig:compEffectOverall}\fbox{\includegraphics[width=4.25cm]{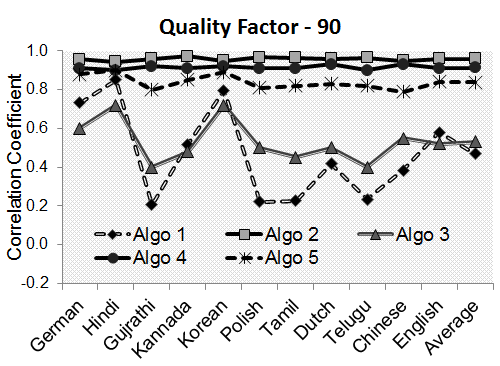}}}%
\caption{Robustness against Compression Attack}
\label{fig:compAttackCorrelation}
\end{figure}

Figure \ref{fig:compEffectChar} shows the effect of compression on characters. From Figure \ref{fig:compEffectChar}, it is clear that decrease in quality factor makes the words and characters blurry and hence, obscures reading. Besides blurriness, compression may destroy the embedded watermark. Figure \ref{fig:compAttackCorrelation} shows the correlation coefficient of the extracted watermark with respect to the embedded watermark at different quality factors. For all documents under consideration, the average value of correlation coefficient across different languages for Algo 1, Algo 2, Algo 3, Algo 4, and Algo 5 are between [0.38, 0.47], [0.96, 0.97], [0.46, 0.53], [0.90, 0.91], and [0.83, 0.84] respectively when we varied quality factor from 10 to 90.

\begin{figure}[t!]
\centering
\subfigure[variance = 0.1]{\makebox[3cm]{\includegraphics[width=0.27\textwidth]{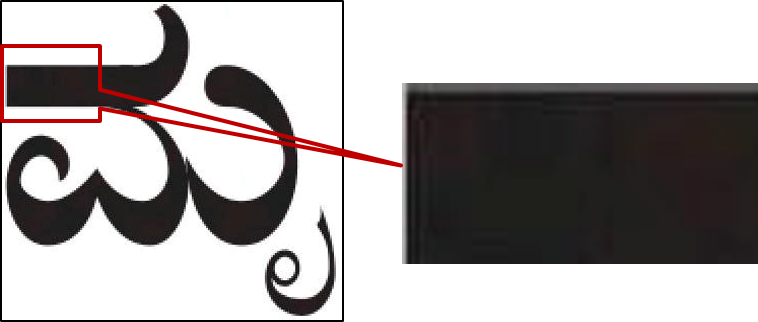}}}
\hspace{0.1\textwidth}
\subfigure[variance = 1]{\makebox[3cm]{\includegraphics[width=0.27\textwidth]{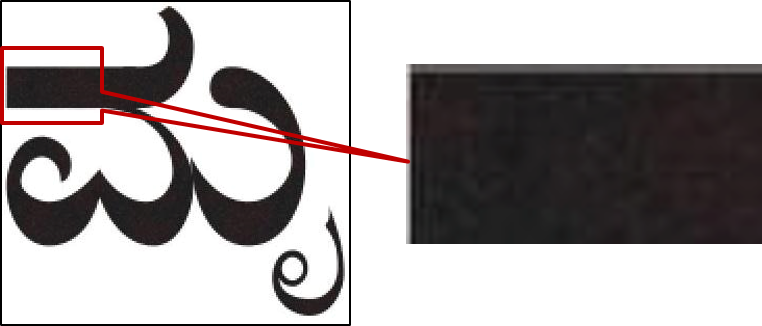}}}
\hspace{0.1\textwidth}
\subfigure[variance = 10]{\makebox[3cm]{\includegraphics[width=0.27\textwidth]{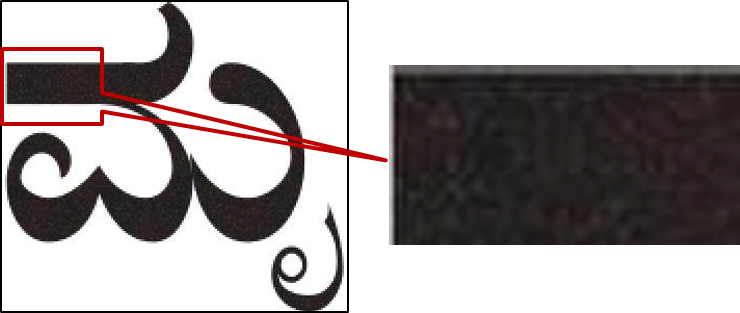}}}
\caption[Effect of Noise on characters]{Effect of Noise on characters (character is from Kannada Language)}
\label{fig:noiseEffectChar}
\end{figure}
\begin{figure}[t!]
\centering
\subfigure[variance = 0.1]{\fbox{\includegraphics[width=4.25cm]{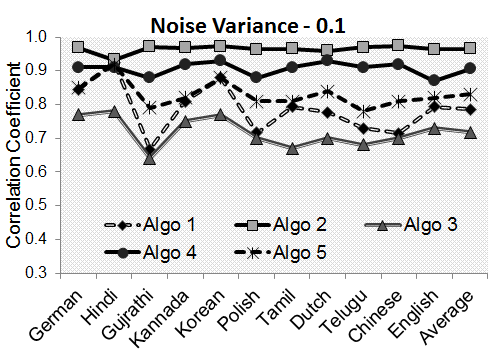}}}%
\vspace{0.001cm}
\subfigure[variance = 1]{\fbox{\includegraphics[width=4.25cm]{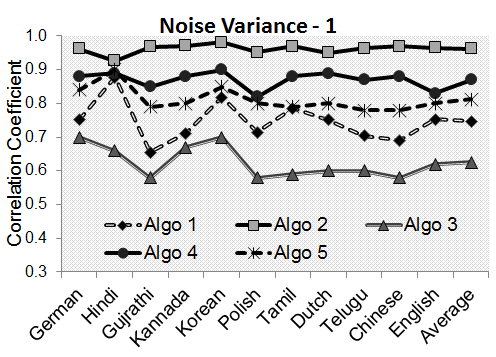}}}%
\vspace{0.001cm}
\subfigure[variance = 10]{\label{fig:noiseEffectOverall}\fbox{\includegraphics[width=4.25cm]{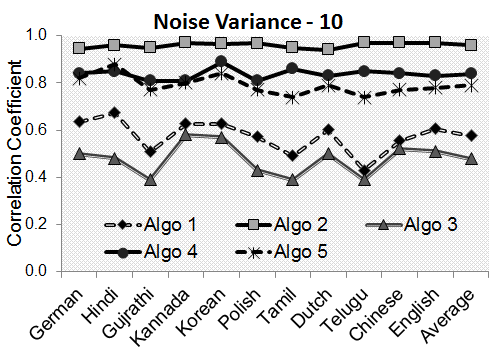}}}%
\caption{Robustness against Noise Attack}
\label{fig:noiseAttackCorrelation}
\end{figure}
\subsubsection{Noise Attack}
\label{sssec:noiseAttack}
Figure \ref{fig:noiseEffectChar} shows the effect of noise addition on characters. From Figure \ref{fig:noiseEffectChar}, it is clear that noise addition does not have a huge impact on the readability of characters. However, addition of noise may destroy the embedded watermark. Figure \ref{fig:noiseAttackCorrelation} shows the correlation coefficient of the extracted watermark with respect to the embedded watermark at different noise levels. For all documents under consideration, the average value of correlation coefficient across different languages for Algo 1, Algo 2, Algo 3, Algo 4, and Algo 5 are between [0.58, 0.79], [0.95, 0.96], [0.48, 0.72], [0.84, 0.91], and [0.79, 0.83] respectively when we varied noise from 0.1 to 10. 

\subsection{Discussion}
The performance of different watermarking algorithms in terms of robustness, computational complexity, and imperceptibility is discussed in above sub-sections. Though all the image-based watermarking algorithms are imperceptible, the robustness against attacks and computational complexity of these algorithms is different. The order of robustness and computational complexity are given in Eqns. \ref{eq:robust} and \ref{eq:complexity} respectively.
\begin{equation}
Algo 2 > Algo 4 > Algo 5 > Algo 1 > Algo 3
\label{eq:robust}
\end{equation}
\begin{equation}
Algo 3 > Algo 1 > Algo 4 > Algo 5 > Algo 2
\label{eq:complexity}
\end{equation}

Algo 2 is more robust then other algorithms. This is due to inherent properties of Algo 2 such as resiliency against noise and scaling \cite{RuizhenLiu2002:eke}. It is worth noting that when we did hybrid watermarking by combining different transform-based algorithms (DCT and DWT) with SVD, the robustness of the hybrid algorithms is increased in comparison to their native implementations. The increase in robustness is due to the inherent properties of SVD.

Another point worth noting is that correlation coefficient for few languages such as Hindi, German, and Korean is greater than other languages. This is due to the fact that the documents in these languages contain more graphical data (for instance, documents in Hindi language have a colored background) than others which increased their robustness against attacks such as compression.

In summary, we can say that Algo 2 is a better option when robustness is a prime concern while Algo 4 is better option when both computational complexity and robustness are prime concerns.

\subsection{Application in fingerprinting}
\label{ssec:fingerprinting}
\begin{figure}[t!]
\begin{minipage}{0.5\textwidth}
\centering
\includegraphics[width=0.8\columnwidth]{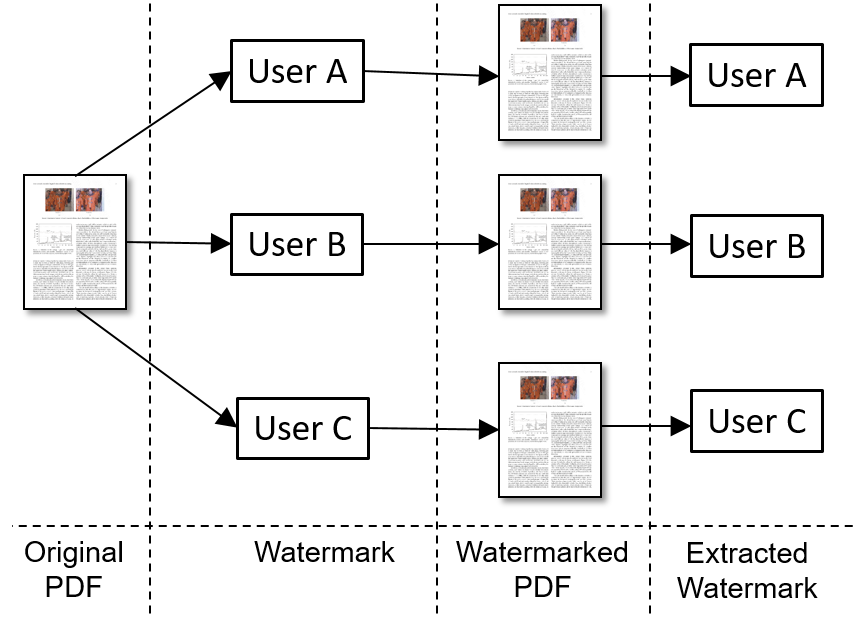}
\caption{Example of fingerprinting}
\label{fig:exampleFingerprint}
\end{minipage}
\hspace{0.5cm}
\begin{minipage}{0.5\textwidth}
\centering
\includegraphics[width=0.8\columnwidth]{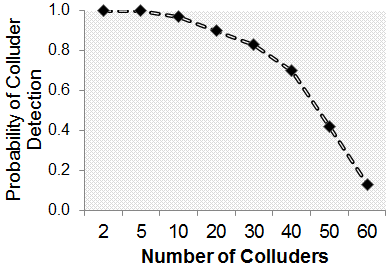}
\caption{Results for Collusion Attack}
\label{fig:collusion}
\end{minipage}
\end{figure}

Fingerprinting aims at embedding a unique watermark for every user. Figure \ref{fig:exampleFingerprint} shows an example of fingerprinting where three instances of watermarked text files are created by embedding three different watermarks. In the proposed method, we generate anti-collusion codes (or unique watermarks or fingerprints) using the method proposed by \cite{Trappe2003:eke} and embed them using Algo 2\footnote{We have chosen the most robust algorithm, however other algorithms can also be used.}. Fingerprints are orthogonal i.e. the correlation between different fingerprints is very less or can be ignored. The generation of anti-collusion codes is out-of-the-scope of this paper and hence, we restrict ourselves from discussing it. However, we encourage readers to read the method proposed by \cite{Trappe2003:eke}. 

Collusion attack is the widely studied attack against watermarking systems aimed at fingerprinting. When a user came with $U$ watermarked copies of same content, he or she can simply average these $U$ copies to create a colluded version of the content. Figure \ref{fig:collusion} shows the result of average collusion attack. From Figure \ref{fig:collusion}, we can see that the proposed method is able to detect at least 90\% of the colluders when number of colluders are less than 20. However, the detection rate drops to $\approx 10\%$ when number of colluders are increased from 20 to 60. The decrease in detection rate is obvious as the quality of colluded content decreases with increase in number of colluders (average PSNR of colluded text documents drop from 55 dB to 14 dB as we increase number of colluders from 2 to 60). 

%==============
\subsection{Comparison with related work}
\label{ssec:compare}
In Table \ref{table:comparison}, we compare the proposed method with state-of-art methods. The capacity of existing watermarking methods is a variable and entirely related to content length and its type. For instance, the embedding capacity of the method proposed by \cite{Low1995:eke} is proportional to number of lines in a document. In contrast to existing methods, the embedding capacity of the proposed method is dependent upon the image watermarking algorithm and block size. 

Further, existing algorithms and the proposed method are robust to print \& scan attack. However, existing algorithms cannot be applied for fingerprinting because malicious users can easily detect and destroy the embedded watermarks by simply comparing the different copies of a same content. Unlike existing methods which embeds the watermark by changing text specific features such as modulation of distance between words and characters, the proposed method uses image-based watermarking algorithm for embedding the watermark. We observed that image-based watermarking algorithms doesn't alter the text specific features and can be used for fingerprinting. 

\begin{table}[t!]
\centering
\caption{Comparison with existing methods}{
\begin{tabulary}{\textwidth}{|J|J|C|C|C|}
\hline
Method & Capacity & \multicolumn{2}{|c|}{Robustness} & Multiple\\
\cline{3-4}
& & Print \& Scan & Fingerprinting & language Support \\
\hline
\cite{Low1995:eke} & approx. equal to no. of lines in a document & Yes & No & Yes \\
\hline
\cite{Kim2003:eke} & approx. equal to no. of words in a document & Yes & No & Yes \\
\hline
\cite{Wenyin2006:eke} & approx. equal to no. of special characters & Yes & No & No \\
\hline
\cite{Topkara2006:eke}& equal to no. of sentence transformations & Yes & No & Yes \\
\hline
\cite{Jixian2012:eke} & depends upon no. of high frequency words in a sentence & Yes & No & Yes \\
\hline
\cite{Halvani2013:eke} & equal to no. of word substitutions & Yes & No & Yes \\
\hline
Proposed method & depends upon image watermarking algorithm & Yes & Yes & Yes \\
\hline
\end{tabulary}}
\label{table:comparison}
\end{table}

%%%%%%%%%%%%%%%%%%%%%%%%%%%%%
\section{Conclusion}
\label{sec:conclusion}
This paper proposes a framework that facilitates the usage of image watermarking algorithms in text documents. The proposed method divides the document into blocks and classifies them into texture and non-texture blocks using energy. The watermark is then embedded inside the texture blocks using content adaptive watermarking strength. For evaluation, the proposed approach is integrated with known image watermarking methods. Experimental results clearly indicate that the proposed method facilitates the usage of known image watermarking algorithms in text documents and is robust attacks.

Besides this, the proposed method is language independent and does not require manual introspection like NLW methods. As the proposed method helps in adopting image watermarking algorithms in text documents, it can be used in different applications, such as fingerprinting, where existing methods cannot be used.

\bibliographystyle{unsrt}
\bibliography{textbib}

\section{Appendix}
\subsection{Histogram-based block classification method}
\label{sec:histApproach}
\begin{figure}[t!]
\centering
\subfigure[CT]{\fbox{\includegraphics[scale=0.3]{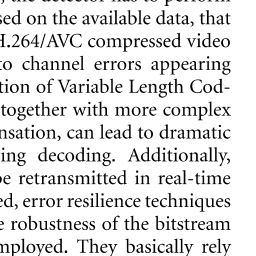}}}
\hspace{0.5cm}
\subfigure[PT]{\fbox{\includegraphics[scale=0.3]{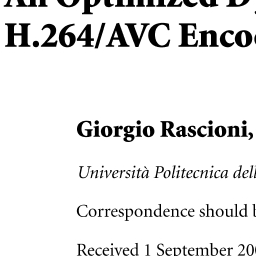}}}
\hspace{0.5cm}
\subfigure[CT]{\fbox{\includegraphics[scale=0.18]{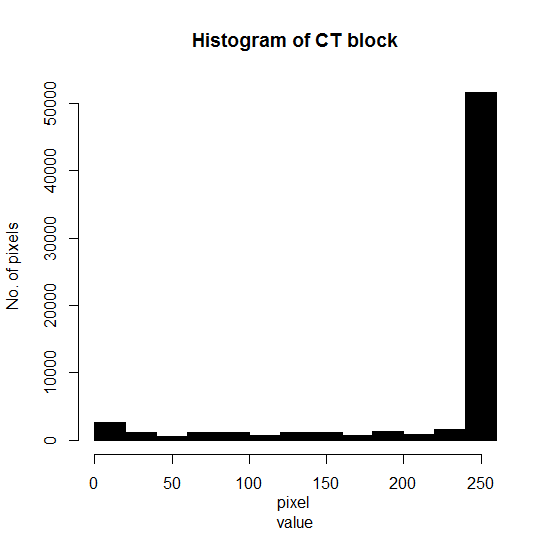}}}
\hspace{0.5cm}
\subfigure[PT]{\fbox{\includegraphics[scale=0.18]{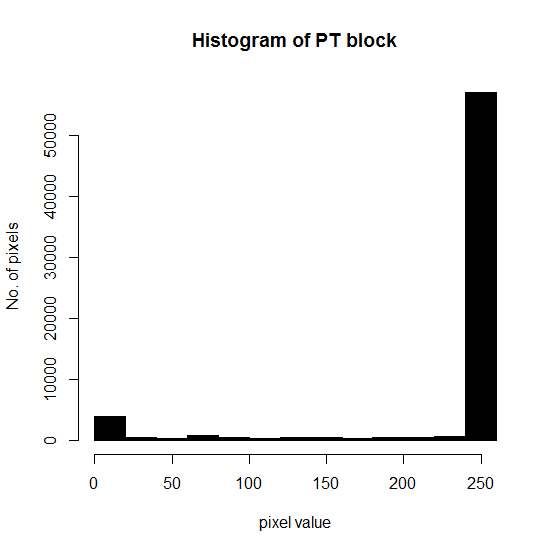}}}
\caption{Histogram of different blocks in a text document: (a-b) Different blocks and (c-d) Histogram of blocks}
\label{fig:hist}
\end{figure}

Histogram captures the information about the distribution of data inside the image. It can be a powerful tool to classify the blocks into two categories i.e. text and graphics. Unlike the energy based approach, it cannot be used for fine grained classification of blocks such as classifying the blocks into CT and PT category. This is because blocks such as CT/PT and CG/PG/PTPG exhibits similar histogram properties and may cause misclassification of the blocks. Figure \ref{fig:hist} shows the histogram of CT and PT blocks. It is evident from the figure that both the blocks exhibit similar histogram properties and hence, fine-grained classification of such blocks is very difficult.

\begin{figure}[t!]
\begin{minipage}{.5\textwidth}
\centering
\includegraphics[width=\columnwidth]{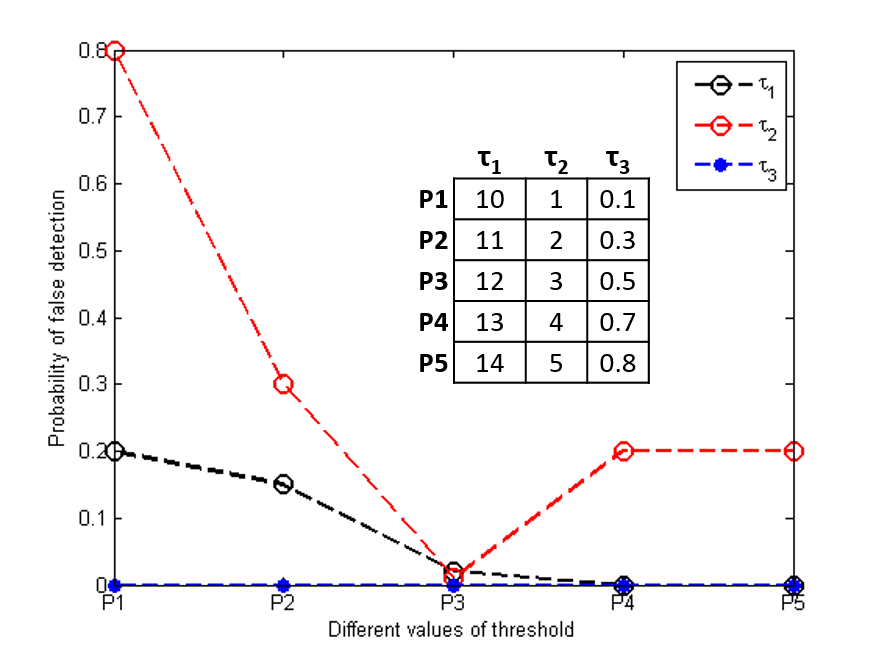}
\caption{Relationship between block classification and thresholds}
\label{fig:blockProbHist}
\end{minipage}
\hspace{0.5cm}
\begin{minipage}{.5\textwidth}
\centering
\includegraphics[width=\columnwidth]{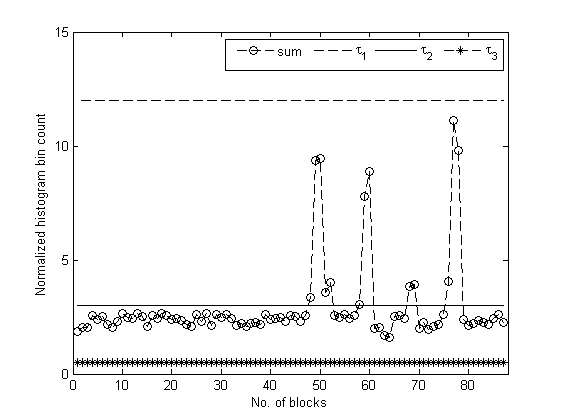}
\caption{Example: Histogram based block classification}
\label{fig:histClassPage}
\end{minipage}
\end{figure}

Block classification using histogram is elaborated in Algorithm \ref{algo:contentClassiHist}. The performance of  Algorithm \ref{algo:contentClassiHist} is dependent upon thresholds ($\tau_1$, $\tau_2$, and $\tau_3$). We conducted several experiments to adjust the value of $\tau_1$, $\tau_2$, and $\tau_3$. Figure \ref{fig:blockProbHist} shows the relationship between the probability of false detection and thresholds. From Figure \ref{fig:blockProbHist}, we can clearly see that probability of false detection is less when thresholds for block classification using histogram ($\tau_1$, $\tau_2$, and $\tau_3$) are 12, 3, and 0.5 respectively. Figure \ref{fig:histClassPage} shows the results for one of the documents used in our experiments. The document contains 88 blocks of dimension $256 \times 256$. From Figure \ref{fig:histClassPage}, it is clear that the histogram based method is able to classify the blocks into text and graphics category.

Apart from energy-based or histogram-based approaches, another possible approach for classification of blocks could be using mean $\mu_b$ and variance $\sigma_b$ of the blocks. As the pixels in text documents take values from a few possibilities and mostly contain black and white pixels, it will be very difficult to classify the blocks using mean $\mu_b$ and variance $\sigma_b$, as shown in Table \ref{table:meanVar}. Hence, we haven't explored this approach. 

\begin{algorithm}[H]
\scriptsize
 \caption{Histogram-based Classification of Blocks}
 Read the block and compute its histogram $H$.
\BlankLine
 Normalize the histogram using L2-normalization as:
\BlankLine
\quad \quad $H_{norm} = \frac{H}{\sqrt{||H||^2} + \epsilon^2}$
\BlankLine
Now, compute the sum of the frequency of colors in $H_{norm}$ as:
\BlankLine
\quad \quad $sum = \sum H_{norm}$
 \BlankLine
 \uIf {$sum > \tau_1$ or $sum < \tau_3$}{
   \emph{Non-texture block}
 }
 \uElse{
    \uIf {$sum \leq \tau_1$ and $sum > \tau_2$}{
       \emph{Graphics block}
    }
    \uElseIf {$sum \leq \tau_2$ and $sum > \tau_3$}{
       \emph{Text block}
    }
}
where $\tau_1$, $\tau_2$, and $\tau_3$ are thresholds used for block classification using histogram.
\label{algo:contentClassiHist}
 \end{algorithm}

\begin{table}[t!]
\centering
\caption{Mean $\mu_b$ and Variance $\sigma_b$ of different blocks in text documents}{
\begin{tabular}{|c|c|c|c|c|}
\hline
\textbf{Block Type} & \textbf{CT} & \textbf{PT} & \textbf{PTPG} & \textbf{CG} \\
\hline
$\mu_b$ & 232.05 & 226.95 & 223.63 & 224.4 \\
\hline
$\sigma_b$ & 12.49 & 19.89 & 18.87 & 14.53 \\
\hline
\end{tabular}}
\label{table:meanVar}
\end{table}

\subsection{Brief overview of image watermarking algorithms}
In this section, we briefly describe the image watermarking algorithms.
\subsubsection{DWT-based watermarking}
In this subsection, DWT-based watermarking algorithm is discussed briefly. A more detailed explanation about the algorithm can be found in \cite{Reddy2005:eke}.

\textbf{Watermark Embedding -} To embed a watermark $W$ inside an image $C$, image $C$ is decomposed into $l$ levels ($C_{l}^{\theta}$) while watermark $W$ is decomposed into $1$ level, i.e. $W_{0}^{\theta}$, using DWT. Here, $l = \{0,1,2,3\}$ represents decomposition level and $\theta\ \epsilon\ \{HH,HL,LH,LL\}$ represents the orientation. Watermark is then embedded as:

\begin{enumerate}[{\bf Step}~1:]
\item First, find the weight factors $w_l^{\theta}$ for wavelet coefficients using the method given by \cite{Barni2001:eke}.
\item Now, add the watermark as: $C_{l}^{' \theta}\ =\ C_{l}^{\theta}\ +\ \beta_a\ w_l^{\theta}\ W_{0}^{\theta}$. Here, $\beta_a$ represents the watermark embedding strength.
\item Watermarked image $C'$ is then obtained by taking the inverse DWT (IDWT) of $C_{l}^{' \theta}$.
\end{enumerate}

\textbf{Watermark Extraction -} To extract the watermark from the possibly attacked image $C''$, following steps are performed:
\begin{enumerate}[{\bf Step}~1:]
\item Watermarked image $C''$ and original image $C$ are decomposed into $l$ levels ($C_{l}^{'' \theta}$) using DWT and then the band coefficients of watermark $W'$ are extracted as:
   \begin{equation}
	 	W_{0}^{' \theta}\ =\ \frac{(C_{l'}^{'' \theta}\ - C_{l'}^{\theta})}{w_l^{\theta}}
		 \label{eq:dwtExtract}
	\end{equation}
\item Now, extracted watermark band coefficients are combined with the distortion parameter as:
	\begin{equation}
		 W_{0}^{'' \theta} = W_{0}^{'' \theta} + W_{0}^{' \theta} \times \left[\frac{w_l^{\theta}\ 2^l}{\sqrt{D_l^\theta}} \right]^2
	\end{equation}
where $D_l^\theta$ is the distortion calculated in the neighbourhood $N_x \times N_y$ as:
 \begin{equation}
D_l^\theta (i,j) =\frac{\sum_{x=i-\frac{N_x}{2}}^{i+\frac{N_x}{2}} \sum_{y=j-\frac{N_y}{2}}^{j+\frac{N_y}{2}} \left[(C_{l'}^{'' \theta}(x,y)\ - C_{l'}^{\theta}(x,y))\right]^2}{N_x \times N_y}
\end{equation}

\item Now, extracted watermark band coefficients are normalized as:
\begin{equation}
 W_{0}^{''\theta} = \frac{W_{0}^{''\theta}}{sum}, where\  sum = sum + \left[\frac{w_l^{\theta}\ 2^l}{\sqrt{D_l^\theta}} \right]^2
\end{equation}
\item IDWT of $W_{0}^{''\theta}$ is taken to form the extracted watermark $W''$.
\end{enumerate}

\subsubsection{SVD based watermarking}
\label{sssec:svdWatermark}
In this subsection, SVD based watermarking algorithm is discussed briefly. A more detailed explanation about the algorithm can be found in \cite{RuizhenLiu2002:eke}.

\textbf{Watermark Embedding:} To embed the watermark $W$ inside the image $C$, we follow below mentioned steps:
\begin{enumerate}[{\bf Step}~1:]
\item Compute SVD of both $C$ and $W$ as: $C\ =\ U_{c}\ S_{c}\ V^T_{c}, \quad W\ =\ U_{w}\ S_{w}\ V^T_{w}$.
\item To embed the watermark inside the cover image, modify the singular values of $C$ as: $S_{W}\ =\ S_{c}\ +\ (\beta_b\ S_{w})$. Here $\beta_b$ is the watermark embedding strength.
\item Now, watermarked image $C'$ can be obtained as: $C'\ =\ U_{c}\ S_{W}\ V^T_{c}$.
\end{enumerate}

\textbf{Watermark Extraction:} Following steps are performed to extract the watermark from the possibly attacked or watermarked image $C''$:
\begin{enumerate}[{\bf Step}~1:]
\item Compute SVD of $C''$ as: $C'' = U_{a} S_{a} V^T_{a}$.
\item Now, extract the singular values of embedded watermark as: $S'_{w} = \frac{(S_{a}-S_{c})}{\beta_b}$.
\item Extracted watermark $W'$ can then be formed as: $W' =  U_{w} S'_{w} V^T_{w}$.
\end{enumerate}

\subsubsection{DCT based Watermarking}
In this subsection, DCT based watermarking algorithm is discussed briefly. A more detailed explanation about the algorithm can be found in \cite{cox:eke}.

\textbf{Watermark Embedding:} To embed the watermark $W$ inside the image $C$, we follow below mentioned steps:
\begin{enumerate}[{\bf Step}~1:]
\item Apply forward DCT to image $C$.
\item Embed the watermark $W$ inside the AC coefficients as: $v_i'\ =\ v_i(1\ +\ \beta_c\ w_i)$, where $v_i$ represents the AC coefficients obtained by applying DCT to image $C$ and $w_i$ is the binary representation of watermark $W$ (During binary conversion, we represent pixel value of 255 as 1 while pixel value of 0 as 0).
\item Now, apply inverse DCT on the modified coefficients to obtain the watermarked image $C'$.
\end{enumerate}

\textbf{Watermark Extraction:} To extract the watermark from the watermarked or possibly attacked image $C''$, we follow below mentioned steps:
\begin{enumerate}[{\bf Step}~1:]
\item Apply forward DCT to $C''$.
\item Extract the binary watermark $w''$ as: $w''_i\ =\ \frac{|v_i''\ -\ v_i|}{\beta_c}$
\item Now, create a watermark image $W''$ from binary watermark $w''$ as:
\begin{equation}
W''_i = \left\{
\begin{array}{cc}
255, & if\ |w''_i|\ >\ tolerance\\
0, & Otherwise
\end{array}
\right.
\end{equation}
\end{enumerate}

\subsubsection{DWT-SVD based Watermarking}
\label{sssec:dwtSVD}
In this subsection, DWT-SVD based watermarking algorithm is discussed briefly. A more detailed explanation about the algorithm can be found in \cite{Ganic2004:eke}.

\textbf{Watermark Embedding:} We follow following steps for embedding the watermark:
\begin{enumerate}[{\bf Step}~1:]
\item Compute l-level and 1-level DWT of image $C$ and watermark $W$ i.e. $C_{l}^{\theta}$ and $W_{0}^{\theta}$ respectively. Here, $l = \{0,1\}$ represents decomposition level and $\theta\ \epsilon\ \{HH,HL,LH,LL\}$ represents the orientation.
\item Apply SVD to each sub-band of $C_{l}^{\theta}$ and $W_{0}^{\theta}$ as: $C_{l}^{\theta} = U_{cl}^{\theta} S_{cl}^{\theta} V_{cl}^{T^{\theta}}, W_{0}^{\theta} = U_w^{\theta} S_w^{\theta} V_w^{T^{\theta}}$.
\item To embed the watermark, modify the singular values of $C_{l}^{\theta}$ as: $S_{cl}^{'\theta} = S_{cl}^{\theta} + \beta_d S_w^{\theta}$. Here $\theta\ \epsilon\ \{HH,HL,LH,LL\}$ represents orientation and $\beta_d$ represents the watermark embedding strength.
\item Now, compute the inverse SVD to obtain the modified sub-bands as: $C_{l}^{'\theta} = U_{cl}^{\theta} S_{cl}^{'\theta} V_{cl}^{T^{\theta}}$.
\item Apply inverse DWT to modified sub-bands $C_{l}^{'\theta}$ to obtain the watermarked image $C'$.
\end{enumerate}

\textbf{Watermark Extraction - } Watermark is extracted from the possibly attacked or watermarked image as:
\begin{enumerate}[{\bf Step}~1:]
\item Compute l-level DWT of image $C''$ i.e. $C_{l}^{''\theta}$.
\item Compute SVD of each sub-band as: $C_{l}^{''\theta} = U_{cl}^{''\theta} S_{cl}^{''\theta} V_{cl}^{''T^{\theta}}$.
\item Now, extract the sub-bands of embedded watermark as: $S_{w}^{'\theta} = \frac{(S_{cl}^{''\theta} - S_{cl}^{\theta})}{\beta_d}$.
\item Apply inverse SVD on extracted sub-bands of watermark as: $W_{0}^{' \theta} = U_w^{\theta} S_{w}^{'\theta} V_w^{T^{\theta}}$.
\item Inverse DWT is now applied to these sub-bands $W_{0}^{' \theta}$ to obtain the watermark $W'$.
\end{enumerate}

\subsubsection{DWT-DCT-SVD based Watermarking}
\label{sssec:dwtSVDDCT}
In this subsection, DWT-DCT-SVD based watermarking algorithm is discussed briefly. A more detailed explanation about the algorithm can be found in \cite{Navas2008:eke}.

\textbf{Watermark Embedding}
We follow following steps for embedding the watermark:
\begin{enumerate}[{\bf Step}~1:]
\item Compute l-level and 1-level DWT of image $C$ and watermark $W$ i.e. $C_{l}^{\theta}$ and $W_{0}^{\theta}$ respectively. Here, $l = \{0,1\}$ represents decomposition level and $\theta\ \epsilon\ \{HH,HL,LH,LL\}$ represents the orientation.
\item Apply DCT to the subbands as:
\begin{equation}
C_{dl}^{\theta}=dct(C_{l}^{\theta})
\end{equation}
\item Now, apply SVD to each sub-band of $C_{dl}^{\theta}$ and $W_{0}^{\theta}$ as:
\begin{equation}
C_{dl}^{\theta} = U_{cl}^{\theta} S_{cl}^{\theta} V_{cl}^{T^{\theta}}, W_{0}^{\theta} = U_w^{\theta} S_w^{\theta} V_w^{T^{\theta}}\\\\
\end{equation}

\item To embed the watermark, modify the singular values of $C_{dl}^{\theta}$ as
\begin{equation}
	S_{cl}^{'\theta} = S_{cl}^{\theta} + \beta_e S_w^{\theta}
\end{equation}
where $\theta\ \epsilon\ \{HH,HL,LH,LL\}$ represents orientation and $\beta_e$ represents the watermark embedding strength.
\item Now, compute the inverse SVD and inverse DCT to obtain the modified sub-bands as:
\begin{equation}
C_{l}^{'\theta}=idct(C_{dl}^{'\theta}) = idct(U_{cl}^{\theta} S_{cl}^{'\theta} V_{cl}^{T^{\theta}})
\end{equation}
\item Now, apply inverse DWT to modified sub-bands $C_{l}^{'\theta}$ to obtain the watermarked image $C'$.
\end{enumerate}

\textbf{Watermark Extraction - } Watermark is extracted from the possibly attacked or watermarked image as:
\begin{enumerate}[{\bf Step}~1:]
\item Compute l-level DWT of image $C''$ i.e. $C_{l}^{''\theta}$.
\item Compute DCT of each sub-band as:
\begin{equation}
C_{dl}^{''\theta}=dct(C_{l}^{''\theta})
\end{equation}
\item Compute SVD of each sub-band as:
\begin{equation}
	C_{dl}^{''\theta} = U_{cl}^{''\theta} S_{cl}^{''\theta} V_{cl}^{''T^{\theta}}\\
\end{equation}
\item Now, extract the sub-bands of embedded watermark as:
\begin{equation}
	S_{w}^{'\theta} = \frac{(S_{cl}^{''\theta} - S_{cl}^{\theta})}{\beta_e}\\
\end{equation}
\item Apply inverse SVD on extracted sub-bands of watermark as:
\begin{equation}
	W_{0}^{' \theta} = U_w^{\theta} S_{w}^{'\theta} V_w^{T^{\theta}}\\
\end{equation}
\item Inverse DWT is now applied to these sub-bands $W_{0}^{' \theta}$ to obtain the watermark $W'$.
\end{enumerate}
\end{document}